\documentclass[aps,prx,reprint,
superscriptaddress,floatfix,longbibliography,footinbib]{revtex4-1}
\usepackage{mathtools}
\usepackage{amsfonts}
\usepackage{amsmath,comment}
\usepackage{tabularx}
\usepackage{physics}
\usepackage{float}
\usepackage{bbold}
\usepackage{xcolor}

\newcommand{\be}{\begin{equation}}
\newcommand{\ee}{\end{equation}}
\newcommand{\bea}{\begin{eqnarray}}
\newcommand{\eea}{\end{eqnarray}}

\begin{document}

\title{Phantom energy in the nonlinear response of a quantum many-body scar state}

\author{Kangning Yang}
\altaffiliation[K.~Y. and Y.~Z.~contributed equally to this work.]{}
\affiliation{Department of Physics, Stanford University, Stanford, CA 94305, USA}
\affiliation{E.~L.~Ginzton Laboratory, Stanford University, Stanford, CA 94305, USA}
\author{Yicheng Zhang}
\altaffiliation[K.~Y. and Y.~Z.~contributed equally to this work.]{}
\affiliation{Homer L.~Dodge Department of Physics and Astronomy, The University of Oklahoma, Norman, OK 73019, USA}
\affiliation{Center for Quantum Research and Technology, The University of Oklahoma, Norman, OK 73019, USA}
\author{Kuan-Yu Li}
\affiliation{E.~L.~Ginzton Laboratory, Stanford University, Stanford, CA 94305, USA}
\affiliation{Department of Applied Physics, Stanford University, Stanford, CA 94305, USA}
\author{Kuan-Yu Lin}
\affiliation{Department of Physics, Stanford University, Stanford, CA 94305, USA}
\affiliation{E.~L.~Ginzton Laboratory, Stanford University, Stanford, CA 94305, USA}
\author{\\Sarang Gopalakrishnan}
\affiliation{Department of Electrical and Computer Engineering, Princeton University, Princeton, NJ 08544, USA}
\author{Marcos Rigol}
\affiliation{Department of Physics, Pennsylvania State University, University Park, PA 16802, USA}
\author{Benjamin L.~Lev}
\affiliation{Department of Physics, Stanford University, Stanford, CA 94305, USA}
\affiliation{E.~L.~Ginzton Laboratory, Stanford University, Stanford, CA 94305, USA}
\affiliation{Department of Applied Physics, Stanford University, Stanford, CA 94305, USA}

\date{\today}

\begin{abstract} 

Quantum many-body scars are notable as nonthermal states that exist at high energies. Here, we use attractively interacting dysprosium gases to create scar states that are stable enough be driven into a strongly nonlinear regime while retaining their character. We uncover an emergent nonlinear many-body phenomenon, the effective transmutation of attractive interactions into repulsive interactions. We measure how the kinetic and total energies evolve after quenching the confining potential. Although the bare interactions are attractive, the low-energy degrees of freedom evolve as if they repel each other: Thus, their kinetic energy paradoxically \emph{decreases} as the gas is compressed. The missing ``phantom'' energy is quantified by benchmarking our experimental results against generalized hydrodynamics calculations. We present evidence that the missing kinetic energy is stored in very high-momentum modes. 

\end{abstract}

\maketitle

Interactions can restructure the low-energy spectra of quantum systems~\cite{Laughlin2000tto}: Thus, the elementary excitations of a system of fermionic atoms can be bosons (if the system goes superfluid) or anyons (if it enters the fractional quantum Hall regime)~\cite{moessner2021tpm}. Traditionally, this notion of emergence was thought to be restricted to ground states. The highly excited states of generic interacting systems are not expected to support long-lived, particle-like excitations: Instead, the degrees of freedom of the system rapidly exchange energy as the system equilibrates. Over the past decade, however, experiments and simulations have shown that some many-body quantum systems can evade thermalization for very long timescales; they can persist in prethermal states either because they are approximately integrable~\cite{Kinoshita2006aqn} or because they possess special ``quantum many-body scar'' eigenstates~\cite{Vafek2017eoe,Bernien2017pmd,Turner2018web,Moudgalya2018ees,Khemani2019soi,Serbyn2021qms,Moudgalya2022qms}.  Scars are expected to be unstable to weak perturbations~\cite{Lin2020sto}. Thus, the response experiments that would normally be used to characterize a stable phase of matter are generally not applicable to scars. While scar linear response has been probed~\cite{Kao2021tpo}, \emph{nonlinear} response experiments that reveal the lifetimes and other emergent properties of elementary excitations would generally be inaccessible as driving a scar out of steady state is expected to make it disintegrate.

\begin{figure}[t!]
    \centering
    \includegraphics[width=1\columnwidth]{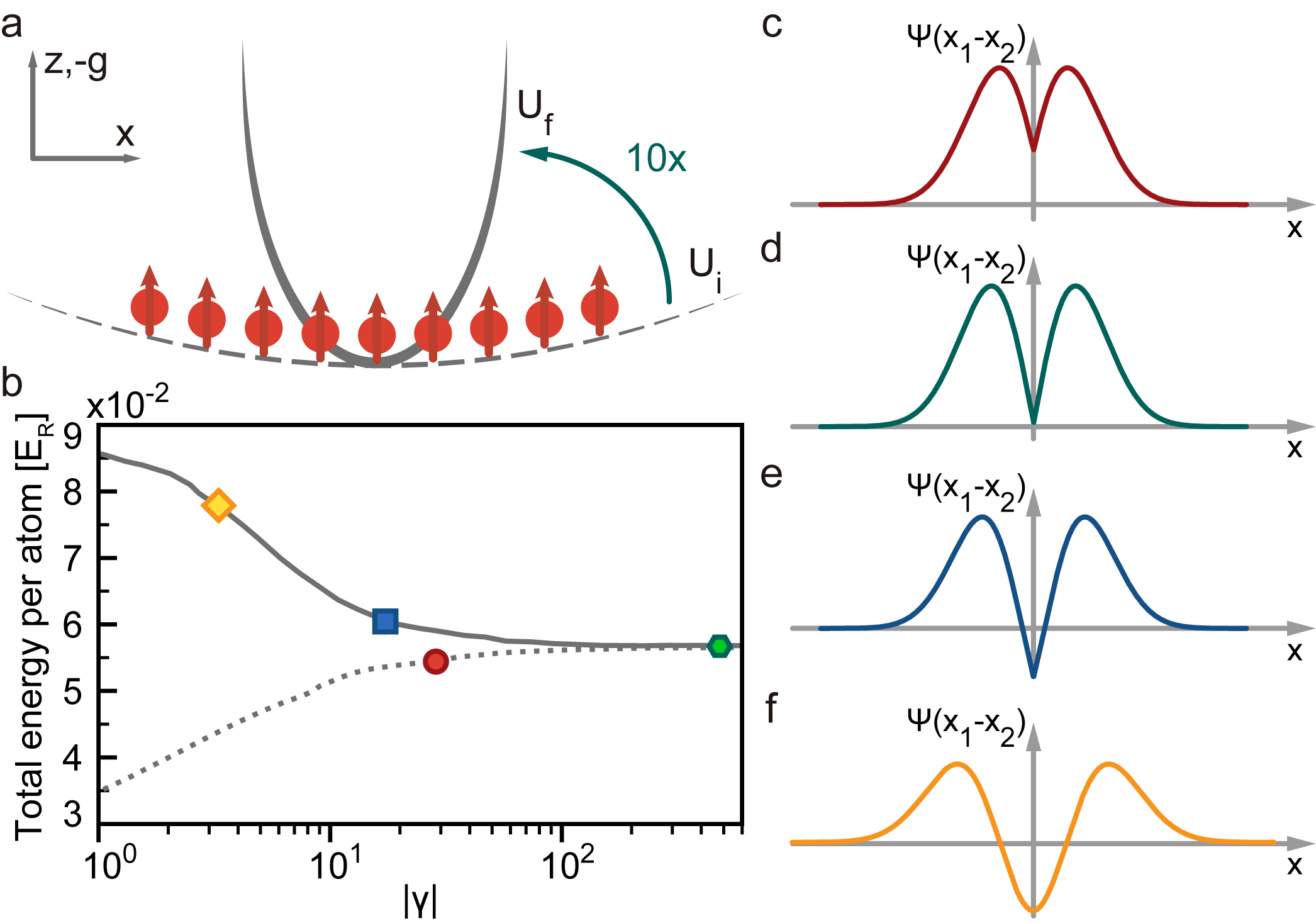}\vspace{-2mm}
    \caption{Trap quench dynamics and states of interest in the first holonomy cycle. (a), Sketch of trap potentials along $\hat{x}$.  The quench is from depth $U_i$ (dashed) to $U_f$ (solid).  Atomic dipoles are aligned along $\hat{z}$. (b), Total energy per particle versus the $|\gamma|$ of the four states studied.  The repulsive (attractive) branch of the first quantum holonomy cycle is shown as a dotted (solid) curve; markers and curves are from the model of the experiment at finite temperature~\cite{Supp}. (c-f), Sketches of the relative wavefunctions of two bosons in a harmonic trap: (c), a repulsive gas at $\gamma=26.3$ [red circle in panel (b)]; (d), the near-unitary sTG gas at $\gamma=-480$ (green hexagon); (e), a scar state at $\gamma = -15.5$ (blue square); and (f), a weakly attractive-interacting excited state at $\gamma = -3.4$ (orange diamond)~\cite{Zurn2012fot}. We will use these sketches in future insets to indicate the state to which data correspond. 
    }\vspace{-5mm}
    \label{fig1}
\end{figure} 

Despite this expectation, in this work we create quantum many-body scars that are sufficiently stable that their nonlinear response can be probed. This allows us to characterize and \emph{quantify} an emergent phenomenon we find in these states: Namely, the transmutation of attractive interactions among the bare particles into emergent repulsive interactions among the low-momentum modes. By quenching the interactions in a one-dimensional (1D) atomic Bose gas from strongly repulsive to strongly attractive, we create a highly excited state---the so-called super-Tonks-Girardeau (sTG) gas~\cite{Astrakharchik2004qbg}---in which the particles avoid each other, like hard rods~\cite{Haller2009roa,Girardeau1960rbs}. The atomic dipole-dipole interaction (DDI) stabilizes the gas so we can explore previously inaccessible regimes beyond this unitary, integrable limit. These host quantum many-body scars, in which particles are strongly correlated~\cite{Kao2021tpo}. Having prepared this strongly interacting prethermal state, we rapidly compress it (Fig.~\ref{fig1}a) and separately measure how the kinetic and total (kinetic + interaction) energy evolve. The evolution of the total energy shows that we are able to prepare highly excited states; see Fig.~\ref{fig1}b. Paradoxically, as the gas compresses and the underlying bosons get closer to one another, the observed kinetic energy of the accessible momentum modes decreases, although the microscopic interactions are purely attractive.    This seems to violate energy conservation since the total kinetic energy must go up as the gas compresses. The process is reversible, and the missing (or ``phantom'') energy reappears when we subsequently let the gas expand. 

To quantify the phantom interaction energy, we need a separate estimate of the \emph{true} kinetic energy of the compressed gas. We are able to compute this using the generalized hydrodynamics (GHD) framework~\cite{CastroAlvaredo2016ehi,Bertini2016tio}, which we benchmark here using quenches of gases in near-ground-state repulsive and highly excited attractive regimes. We provide evidence that the phantom energy is stored in the high-energy tails of the bosonic momentum distribution function, resolving the apparent violation. That is, focusing on the low-energy/long-wavelength modes leads to an emergent repulsive interaction that seems to violate conservation of energy. The contradiction is resolved by noting that the high-energy/short-wavelength modes compensate for the apparent loss of energy. 

The origin of the phantom energy can be understood as follows. Consider two particles in a box of size $L$, interacting via an attractive short-range potential. In the ground state, the two particles are bound; in excited states, they avoid each other (because excited eigenstates must remain orthogonal to the bound state). Decreasing $L$ reduces the available space, so the energy of the excited states goes up, as if the system had repulsive interactions. The strength of these apparent repulsive interactions can be estimated from the suppression of the probability for two particles to be near one another in real space. However, in practice, this estimate falls short of the total kinetic energy because it misses the short-distance structure due to hard-to-resolve wavefunction wiggles. In momentum space, this manifests as momentum tails that cannot be detected.   Remarkably, although the system we are studying is a strongly interacting many-body system, this intuition qualitatively captures many of the phenomena we observe. The evolution of the two-particle wavefunction as one tunes the interactions from repulsive to attractive is shown in Fig.~\ref{fig1}c-f.

\begin{figure}[t!]
\centering
\includegraphics[width=0.95\columnwidth]{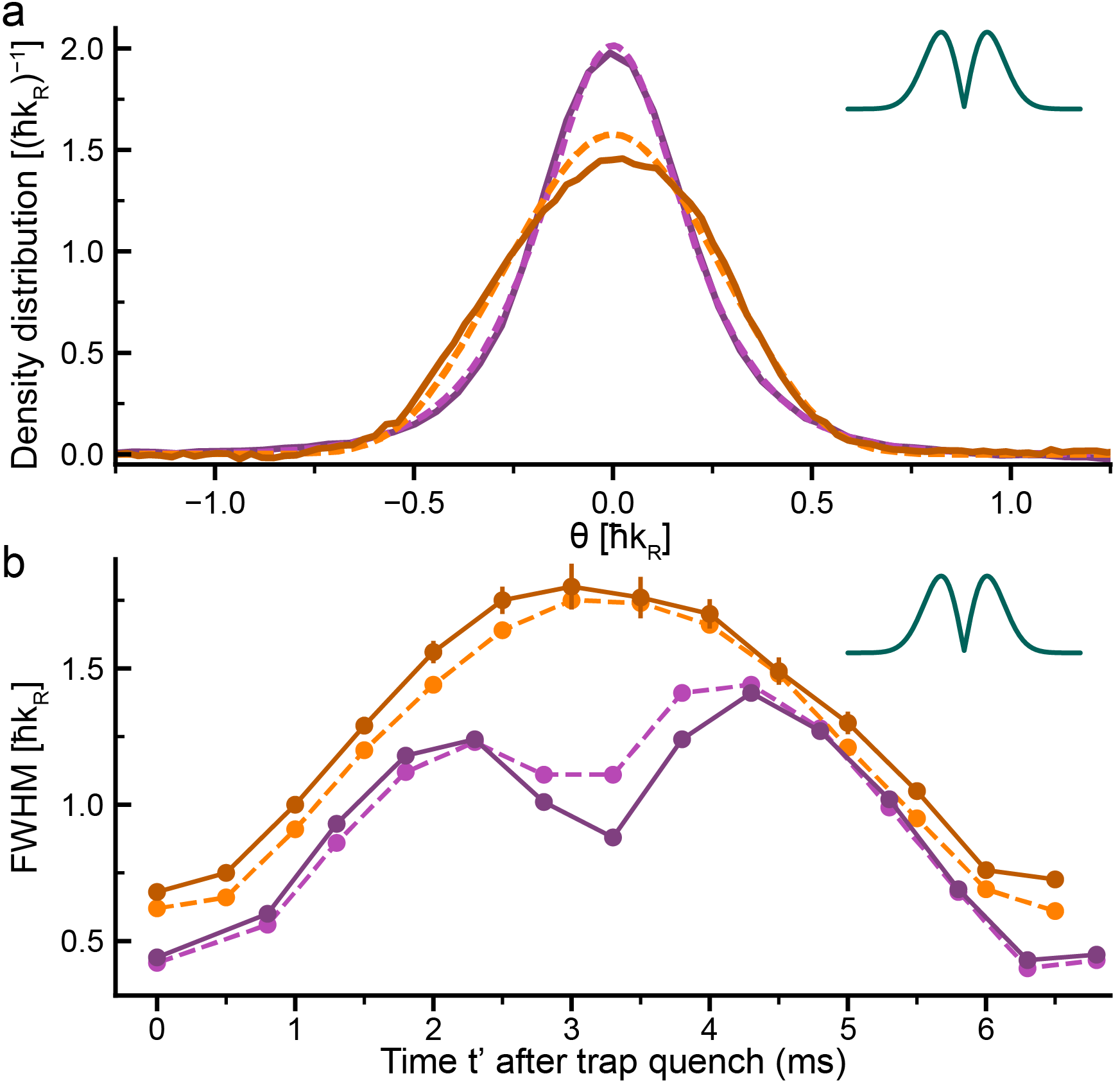}
\caption{Benchmarking the quench experiment in the fully simulatable sTG regime. Solid (dashed) lines represent experimental (simulation) results in both panels. (a), Distributions of momentum (purple) and rapidity (orange) at $t'=0$ for the sTG gas.  Here and in Fig.~\ref{fig4}, $\theta$ stands for either momentum or rapidity along $\hat{x}$. (b), Dynamics during the first oscillation period after the 10$\times$-trap quench. The purple (orange) trace shows the time evolution of the momentum (rapidity) distribution's FWHM.  Data in panel (b) are the average of $\sim$60 shots. $\hbar k_\mathrm{R}$ is the recoil momentum at 741~nm. Theory calculations are done in the TG limit~\cite{Supp}.  Error bars in (b) and subsequent figures are explained in Ref.~\cite{Supp}.}\label{fig2}\vspace{-5mm}
\end{figure}

\begin{figure*}[t!]
    \centering
    \includegraphics[width=0.86\textwidth]{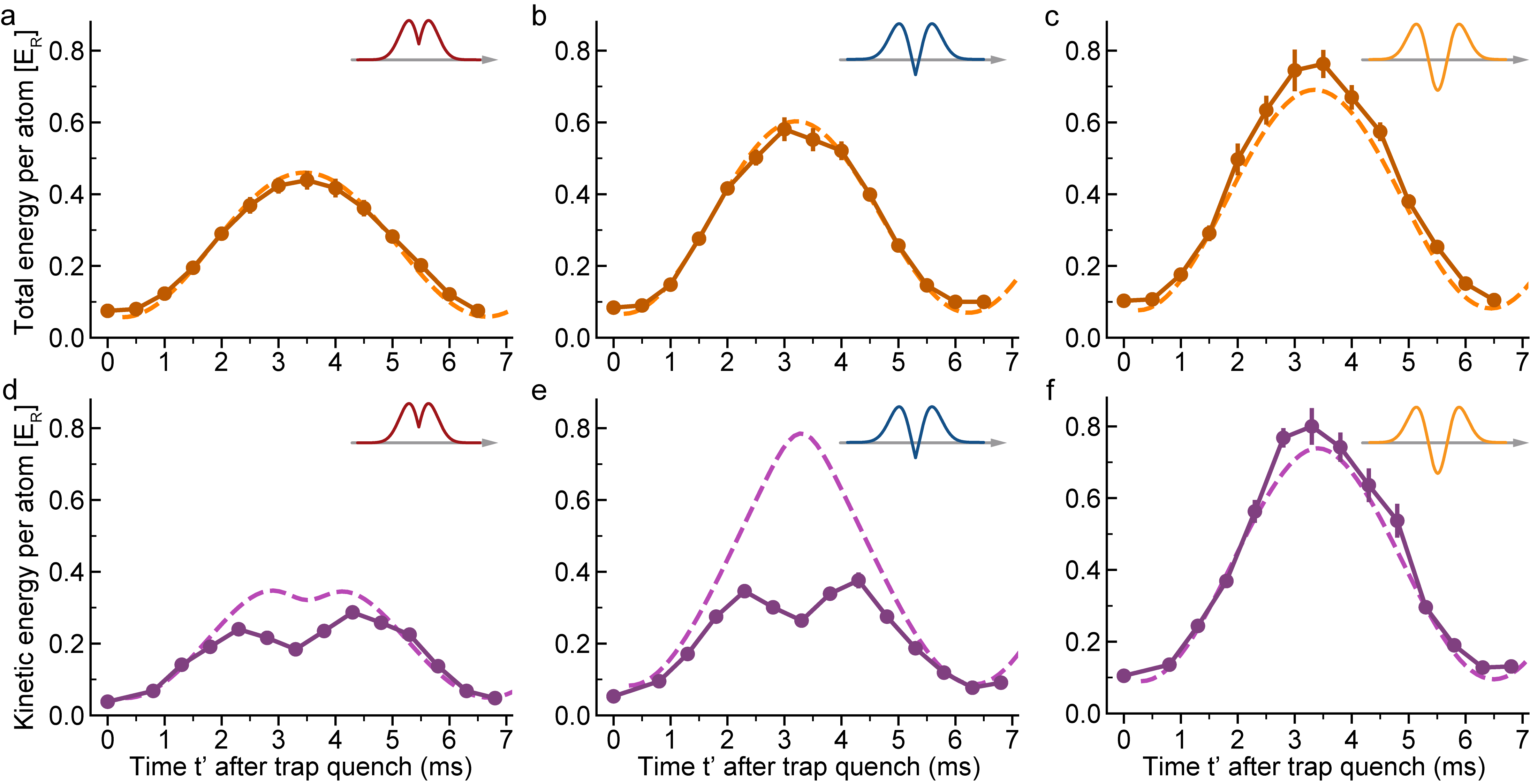}
\caption{Evidence for missing energy. The quench dynamics of the repulsive state in panels~(a,d) is compared to that of the scar state (b,e) and noninteracting excited state (c,f). (a-c), Total energy per atom obtained from rapidity distributions. Experimental data (solid) are compared to GHD simulations (dashed) during the first oscillation period. Panels (d-f) show the same for the kinetic energy per atom obtained from momentum distributions. Theory and experiment are synchronized to correct for finite TOF effects~\cite{Supp}. The recoil energy is $E_\mathrm{R}= \hbar^2 k_\mathrm{R}^2/2m$. Data points are averaged $\sim$60 times.}
    \label{fig3}
\end{figure*} 

The excited states are experimentally accessible via a topological pumping method~\cite{Kao2021tpo}.  The pump exploits the quantum holonomy inherent to 1D bosonic gases describable by the Lieb-Liniger Hamiltonian~\cite{Yonezawa2013qhi,Lieb1963eao_2}: 
\be
\label{eq:H_LLmain}
H=\sum_{i=1}^{N}\left[-\frac{\hbar^2}{2m}\frac{\partial^2}{\partial x_i^2}+U_{\rm H}(x_i)\right]+\sum_{1\leq i<j 
\leq N}g_{\rm 1D}\delta(x_i-x_j).
\ee
We added a harmonic confining potential $U_{\rm H}$ and defined $N$ as the number of atoms and $m$ the atomic mass. The pump works by tuning the effective 1D contact interaction strength $g_{\rm 1D}$ in a cycle from $+0$, to $+\infty$, through a quantum quench to $-\infty$, and back to $-0$. While the Hamiltonian returns to itself at the end of the holonomic cycle, the ground state is pumped to an excited energy eigenstate. Dipolar stabilization of the highly magnetic gas allows the system to remain stable throughout the cycle~\cite{Kao2021tpo,Chen2023otm}. While the long-range portion of the 1D DDI is not captured in Eq.~\eqref{eq:H_LLmain}, we incorporate the leading-order, short-range component by adding it to the van der Waals contact strength: $g_{\rm 1D}=g_{\rm 1D}^{\rm vdW}+g_{\rm 1D}^{\rm DDI}$~\cite{Kao2021tpo,Li2023ram,Supp}.

Our experiments will compare properties of the quenched quantum many-body scar state to those of three other states found along the cycle. For concision, we denote these four states as the repulsive, sTG, scar, and weakly attractive states, as explained below. The measurements involve observations of their rapidity and momentum distributions. ``Rapidities'' are the generalized momenta of a set of emergent stable quasiparticles and therefore include the influence of both their interparticle interactions and kinetic energy.  Integrating the square of the rapidity using the rapidity distribution provides the total energy, while doing the same with the momentum using its distribution provides only the kinetic energy.

We briefly discuss the experimental system; see Refs.~\cite{Li2023ram,Kao2021tpo,Supp} for more details.  A 3D BEC of $^{162}$Dy is transferred into a 2D optical lattice that forms $\sim$1500 parallel 1D traps filled with up to 20 atoms in each.  Atoms in each tube are confined within the quasi-1D limit in the $\hat{y}$-$\hat{z}$ plane. The lattice beams plus a crossed optical dipole trap (ODT) provide a weak longitudinal harmonic potential along $\hat{x}$; see Fig.~\ref{fig1}a. A magnetic $B$ field is imposed to polarize the magnetic dipoles along $\hat{z}$, yielding a repulsive intratube DDI.  We tune $g_{\rm 1D}$ by sweeping $|B|$ through a 1D collisional resonance~\cite{Haller2010cri,Kao2021tpo,Supp}.

The trap quench begins with a $10\times$-compression of its depth along $\hat{x}$ by jumping the power of an ODT beam. The gas is allowed to evolve in the longitudinally compressed trap for a variable time $t'$ within one oscillation period.  Atom loss is negligible in the first period~\cite{Supp}, which is shorter than the thermalization time of the states~\cite{Kao2021tpo}. To measure momentum, all trapping fields are turned off at $t'$, allowing the gas to expand in 3D. The distribution of momentum along $\hat{x}$, averaged over the tube ensemble, is observed through time-of-flight (TOF) absorption-imaging~\cite{Bloch2008mpw,Supp}. Rapidity distributions are measured by first releasing the gas at $t'$ along a flat 1D trap in $\hat{x}$~\cite{Supp}. After 10~ms, the gas is released from all traps to perform TOF imaging. The 1D expansion allows the atoms to convert their interacting energy into kinetic energy. The subsequent TOF image reveals the rapidity distribution~\cite{Wilson2020ood}; measurements here are the first of attractive-branch states in the holonomy cycle.
 
Figure~\ref{fig1}b shows the total energy per particle obtained from our model of the experimental system at finite temperature~\cite{Supp}. The cycle has two branches, a repulsive, ground-state branch where $\gamma > 0$ and an attractive, excited branch for $\gamma < 0$. The Lieb-Liniger parameter $\gamma\equiv mg_{\rm 1D}/n_{\rm 1D}\hbar^2$ is the normalized interaction strength, where $n_{\rm 1D}$ is the 1D particle density. The branches meet at the anholonomic point $\gamma = \pm\infty$. Starting near zero energy and pumping upwards, we encounter the repulsive state at $\gamma=26.3(7)$ (red), the unitary sTG state at $\gamma=-480(77)$ (green), the scar regime at $\gamma = -15.5(9)$ (blue), and finally the weakly attractive-interacting excited state at $\gamma=-3.4(1.1)$ (orange). Pumping through the anholonomic point converts the ground state into an excited state in which bosons repel each other like hard rods. 

\begin{figure*}[t!]
\centering
\includegraphics[width=1\textwidth]{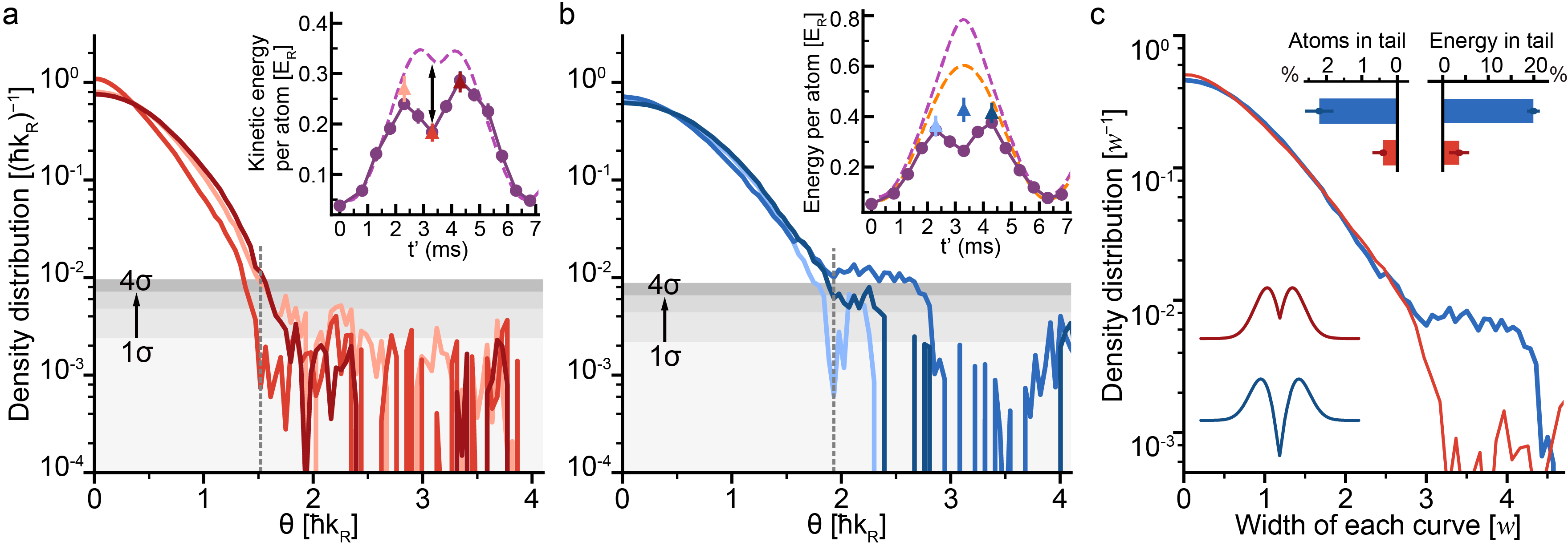}
    \caption{High-resolution momentum distribution measurements. (a,b), Momentum density distribution at three different post-quench times for the (a), repulsive (average of 500 shots) and (b), scar (800) states. In each panel, the three color shades show the momentum distribution at $t' = 2.3$~ms (lightest), 3.3~ms (medium), and 4.3~ms (darkest shade).  The momentum distribution at $t' = 0$~ms (not shown) is used as the residual background reference and is subtracted each curve beyond momenta $\theta \agt 1$~\cite{Supp}. Negative and positive momenta are averaged to make one-sided plots. The gray horizontal color bands mark the statistical uncertainty from the mean background noise in steps of one-to-four standard error ($\sigma$)~\cite{Supp}.    Data at $t'=0$ for all three states are in Ref.~\cite{Supp}. Insets: Time evolution of kinetic energy reproduced from Figs.~\ref{fig3}(d,e). Double-headed arrow indicates kinetic energy difference between theory and experiment near maximum compression. Panel (b) inset also includes the theory curve for total energy (dashed orange line).  Added to these insets are triangular data makers that incorporate the kinetic energy from the high-$k$ tails.  These tails are the data beyond the gray vertical dashed line in each main panel. Their color shade matches the corresponding data in the main panels. (c), Rescaled momentum distribution for $t' = 3.3$~ms for the repulsive (red) and scar (blue) states. Each curve is normalized by the standard deviation of its own width $w$.  Inset: Atom number (left) and kinetic energy (right) at $t' = 3.3$~ms in the momentum distribution tails beyond the high-$k$ cutoffs indicated in panels (a) and (b).   }
    \label{fig4}
\end{figure*} 

Before exploring how this manifests experimentally, we check the reliability of measuring rapidity and momentum for quenched (dipolar) states on the excited, attractive branch; this has already been established for repulsive-branch states~\cite{Wilson2020ood,Malvania2021ghi,Li2023ram}. We focus on the sTG state, which can be modeled using the Tonks-Girardeau (TG) regime of the Lieb-Liniger model~\cite{Astrakharchik2004qbg,Supp}. Figure~\ref{fig2}a shows the sTG rapidity and momentum distributions at $t'=0$, along with exact numerical simulations in the TG regime~\cite{Supp,Li2023ram,Xu2017eoo}. The level of agreement is on par with observations of equilibrium dipolar TG gases~\cite{Li2023ram}. Likewise, Fig.~\ref{fig2}b shows that theory tracks the post-quench evolution of the distribution's full width at half maximum (FWHM). We observe the expected broadening of the rapidity distribution as the gas compresses, as well as a narrowing of the momentum distribution about the maximal compression point due to the transient bosonization of the otherwise fermionized distribution~\cite{Wilson2020ood,Supp}. 

The asymmetry in the FWHM of the momentum distribution about the maximal compression point is a consequence of the finite TOF~\cite{Malvania2021ghi} and is captured by the model~\cite{Supp}. The excellent agreement of all these data validates the measurement technique for the attractive branch.  Moreover, energy conservation, assumed in the simulation, is apparent in the experimental results during the first oscillation. Because atom loss and magnetic energy from either the long-range intratube or intertube DDI are not accounted for in simulation, the close correspondence implies that they play little role in the experimental quench dynamics in the sTG state. (The intertube DDI is estimated to be insignificant~\cite{Supp}.) 

Figure~\ref{fig3} presents quench dynamics of the scar state.  To contextualize our expectations, we first describe the numerical simulations of the repulsive, scar, and weakly attractive state dynamics.  The repulsive state plays the foil because it has a similar interaction magnitude $|\gamma|$ as the scar, but far weaker interatomic correlations. Because the momentum distribution cannot be simulated for these states, we plot either total or kinetic energy per atom rather than FWHM.  As expected, the trap energy decreases so the total (kinetic + interaction) energy increases as the gases reach maximum compression near 3.3~ms.  We also see that the scar and weakly attractive states, lying higher on the holonomy cycle, have a total energy exceeding the repulsive state. The kinetic energy of the repulsive state dips due to the increase in positive interaction energy upon compression~\cite{Malvania2021ghi}.  The scar state exhibits the opposite behavior because its interactions are attractive: To conserve total energy, kinetic energy must \textit{peak} above total energy to compensate the increase in \textit{negative} interaction energy. This is also true for the weakly attractive state, but to a lesser degree.

The experimental total energy in Figs.~\ref{fig3}a-c are well-described by the numerical simulations the rapidity distributions; see Ref.~\cite{Supp} for more comparisons.  This indicates that, for the first time, scar states far from equilibrium can be fully characterized. Moreover, kinetic energies of the repulsive and weakly attractive states also behave as expected, as shown in Figs.~\ref{fig3}d and f, resp. Surprisingly, the kinetic energy of the scar state in Fig.~\ref{fig3}e dips rather than peaks at maximum compression. It is as if the state is \textit{actually} strongly repulsive. This is the key observation of this work.  If energy is conserved, then where does the missing kinetic energy go?  Like a phantom, the missing energy must be a figment of our imagination---it exists but lies outside our field of observation. 

We hunt for the missing energy by averaging additional shots to reduce the high-$k$ noise floor.  Lower noise reveals more of a momentum tail for the scar state than the repulsive; see Fig.~\ref{fig4}. For both states, the peak of the momentum distribution at the maximum compression time of 3.3~ms is taller and narrower than those at 2 and 4~ms. But unlike the repulsive state, the 3-ms data of the scar exhibits a significant momentum tail:  Atomic population exceeds the 4$\sigma$-level out to a momentum of $\theta \approx 2.6\hbar k_\mathrm{R}$ (where it abruptly drops for an unknown reason). This difference is very clear from comparing their width-normalized distributions at 3.3~ms in Fig.~\ref{fig4}c; the repulsive-state distribution is more peaked, while the scar state has a long momentum shoulder. The tail of the integrable sTG state is intermediate between these cases because of the aforementioned bosonization at high density~\cite{Wilson2020ood}; see Ref.~\cite{Supp} for sTG data. 

The data show that the scar state shoulder holds $\sim$2\% of its atoms, which contribute 20\% of the kinetic energy. While this reveals some of where scar's kinetic energy has gone, it does not account for all that is missing. Including the contribution from the new high-$k$ tails does not significantly change the kinetic energy per atom for the repulsive state, but it does completely fill in the scar-state's dip. See the triangular data in the insets of Figs.~\ref{fig4}a,b. However, the new peak in the scar data does not extend past the total energy, let alone the kinetic energy prediction---it still behaves like a repulsive gas.  

Some of the residual missing energy may be due to the long-range DDI. But this likely amounts to no more than a tenth of an $E_\mathrm{R}$ out of the missing ${\sim}0.4E_\mathrm{R}$ at 3.3~ms: The kinetic energy of the repulsive state, being the most dense, bounds the maximum contribution of the long-range DDI. The black arrow in Fig.~\ref{fig4}a shows this to be at most ${\sim}0.1E_\mathrm{R}$.  But the larger shortfall of the scar state cannot be solely due to the DDI. Because the scar state is undoubtedly microscopically attractive, we are left to conclude that its much larger energy deficit is from atoms at momenta beyond what we observe. Indeed, only 240 scar-state atoms would need to exist at, say, $4\hbar k _\mathrm{R}$ to account for all the missing kinetic energy; this is below the current detection limit.  We note that the missing energy does reappear at the decompression end of the oscillation. We also observe---see Ref.~\cite{Supp}---that quenching the trap by only $2\times$ does not produce a dip in the scar state's kinetic energy. The phantom energy seems to be both a coherent and nonlinear effect.   

This scar state, which has attractive interactions at the microscopic level, exhibits emergent repulsive interactions among the low-momentum degrees of freedom. While this phenomenon can be intuitively understood in the context of a two-particle excited state, the surprise is that it persists in the nonlinearly driven, strongly interacting regime of a many-body system, which is far from being in a single excited eigenstate.  The significance of the phantom energy is far broader than a surprising nonlinear response phenomenon: The experiment establishes that driven, highly excited quantum matter can organize in ways distinct from ground and thermal states.

We thank Zhengdong Zhang and Alex Kiral for technical assistance.  We acknowledge the NSF (PHY-2308540) and AFOSR (FA9550-22-1-0366) for funding support. K.-Y.~Lin acknowledges partial support from the Olympiad Scholarship from the Taiwan Ministry of Education. Y.Z.~and M.R.~acknowledge support from the NSF Grant No.~PHY-2012145. S.G.~acknowledges funding from an NSF Career Award.

\onecolumngrid
\clearpage

\section*{Supplementary Information}

\tableofcontents

\setcounter{figure}{0}
\setcounter{equation}{0} 
\renewcommand{\thefigure}{S\arabic{figure}}
\renewcommand{\theequation}{S\arabic{equation}}

\section{Experiment}

\subsection{BEC production, quasi-1D confinement, trap quench, and trap release}

We produce a $^{162}$Dy Bose-Einstein condensate (BEC) in a 1064-nm crossed optical dipole trap (ODT) following the procedure of Ref.~\cite{Li2023ram}. During the evaporation, we ramp the magnitude of bias magnetic field in 1~ms to 26.69~G while keeping the field direction fixed along $\hat{z}$ for better stability.  The field magnitude is experimentally chosen for optimal BEC production close to the 27-G Feshbach resonance:  This allows us to use the confinement induced resonance (CIR)~\cite{Haller2010cri} characterized in Ref.~\cite{Kao2021tpo}. (This is a different resonance than used in Ref.~\cite{Li2023ram}; see Sec.~\ref{sec:FR} for more information about the Feshbach resonance.) The typical atom number is $1.1(2) \times 10^4$ at the end of the evaporation sequence.  The final ODT trap frequency is $2\pi{\times}[55(1), 22.4(5), 113.9(1.5)]$~Hz.

While keeping the 1064-nm crossed ODT on after evaporation, the BEC is loaded into a 2D optical lattice.  This strongly confines the atoms in $\hat{y}$ and $\hat{z}$, forming an ensemble of quasi-1D tube-like traps along $\hat{x}$. As in Ref.~\cite{Li2023ram}, the 2D lattice beams are 5-GHz blue-detuned from the $\lambda = 741$~nm atomic transition. We ramp the lattice to $V_0 = 30 E_R$ in 200~ms, where $E_R/\hbar = 2\pi \times 2.24$~kHz is the recoil energy of a lattice photon. The corresponding transverse trap frequency is $\omega_\perp = 2\pi \times 25$~kHz, with around 20 atoms in the center tube.

After the lattice is fully turned on, we adjust the longitudinal trap shape in 150~ms by ramping down the power of the 1064-nm crossed ODT and ramping up the power of a 1560-nm ODT along $\hat{y}$.  This is used to create a 1D flat trap for the rapidity measurements. Since the blue-detuned lattice forms a longitudinal antitrap of around 7~Hz at a lattice depth of $30 E_R$, the power of the 1560-nm ODT is set such that the antitrapping potential is balanced and forms a flat trap of length 60~$\mu$m along $\hat{x}$. The final trap configuration thus consists of the blue-detuned 2D optical lattice and the red-detuned 1560-nm ODT and 1064-nm crossed ODT. The main contribution to the longitudinal confinement comes from the 1064-nm ODT; we adjust its power such that the overall trap frequency is $\omega_\parallel = 2\pi \times 27.8(5)$~Hz.

We quench the longitudinal trap potential by jumping the power of the 1064-nm ODT within 50~$\mu$s. The quenched longitudinal trap frequency is $\omega_\parallel = 2\pi \times 80.3(8)$~Hz  ($2\pi \times 39.8(3)$~Hz) for the 10$\times$ (2$\times$) quench measurements.  The quench time is randomized for each repetition of the experiment to avoid systematic bias due to drifts.

To implement the 1D expansion, we quickly turn off the 1064-nm ODT to switch off the longitudinal confinement. To map all the quasiparticle interaction energies onto particle momenta, it is critical to allow a $t_{1D} = 10$~ms 1D expansion time before the 16-ms 3D time-of-flight (TOF). As in Ref.~\cite{Li2023ram}, this is chosen to allow the lineshape to stop evolving without reducing the subsequent signal-to-noise ratio (SNR) of the TOF absorption image.  Performing TOF imaging requires deloading the lattice before releasing the atoms to free fall. Deloading takes 300~$\mu$s, resulting in a shift between the rapidity and momentum time evolution measurements. This deload time is short compared to the axial trap period, and is subtracted from the data in the figures.  For the first 100~$\mu$s of the deloading, the lattice beams are kept on while we rapidly adjust the $\hat{z}$ bias magnetic field to where the $s$-wave scattering length is zero.  Then we ramp down the lattice beams in 200~$\mu$s. Little of the van der Waals contact interaction energy is converted into kinetic energy.

\begin{figure*}[t!]
    \centering
    \includegraphics[width=0.9\textwidth]{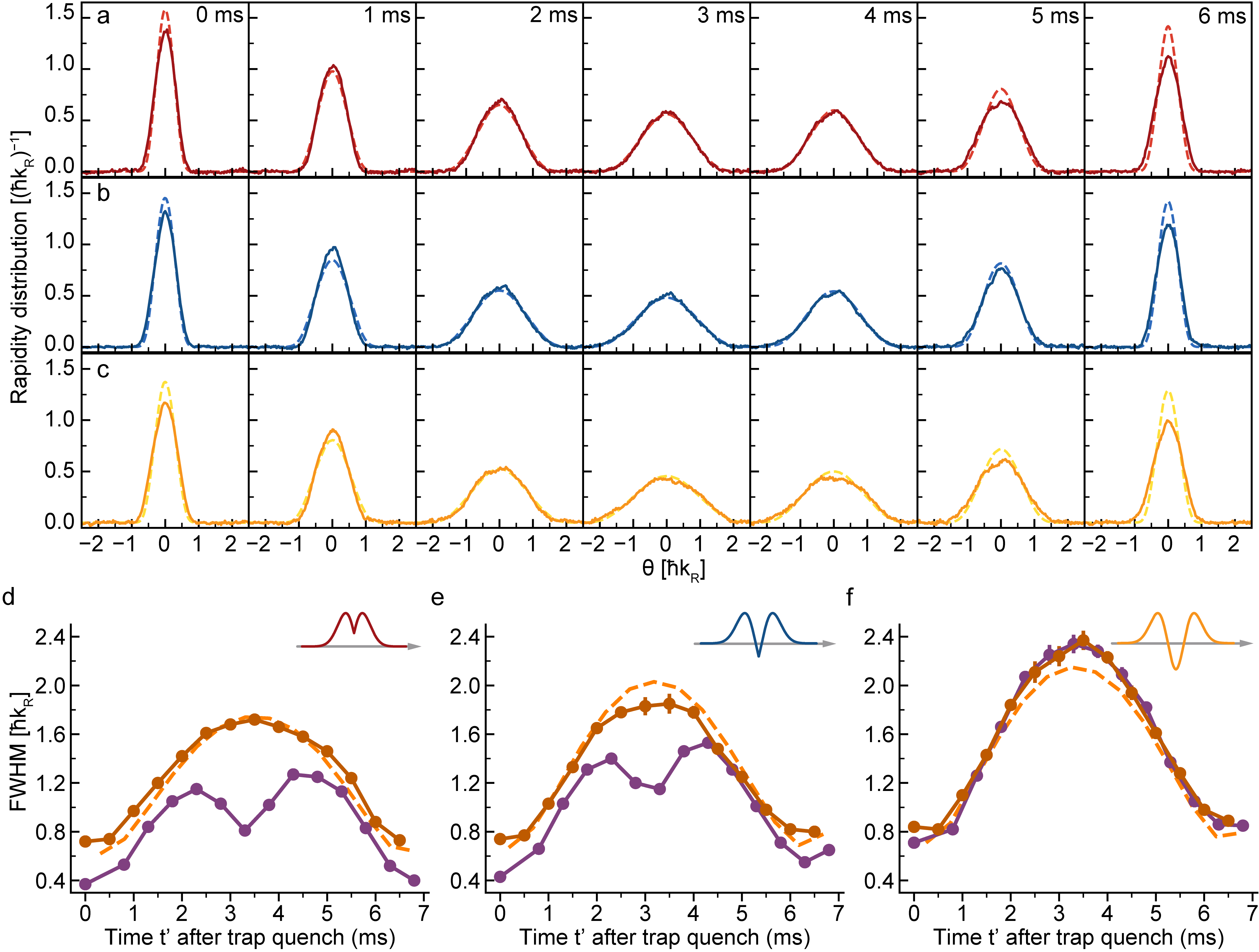}
    \caption{Time evolution of the rapidity and momentum distributions.  (a--c) Direct comparison of rapidity distribution between GHD simulation (dashed) and experiment (solid) for (a), repulsive; (b), scar; and (c) weakly attractive states.  (d--f) Time evolution of FWHM obtained from momentum (purple) and rapidity (solid orange) measurements for (d), repulsive; (e) scar; and (f) weakly attractive state. GHD simulation results for the rapidity FWHM are shown as dashed orange curves. Theory and experiment are synchronized by the same amount as in Fig.~3 to correct for finite TOF effects; the shift times are listed in Fig.~\ref{fig:Parameters} and discussed in Sec.~\ref{sec:TGLimit}. }
    \label{fig:FWHM}
\end{figure*} 

\begin{figure*}
    \centering
    \includegraphics[width=1\textwidth]{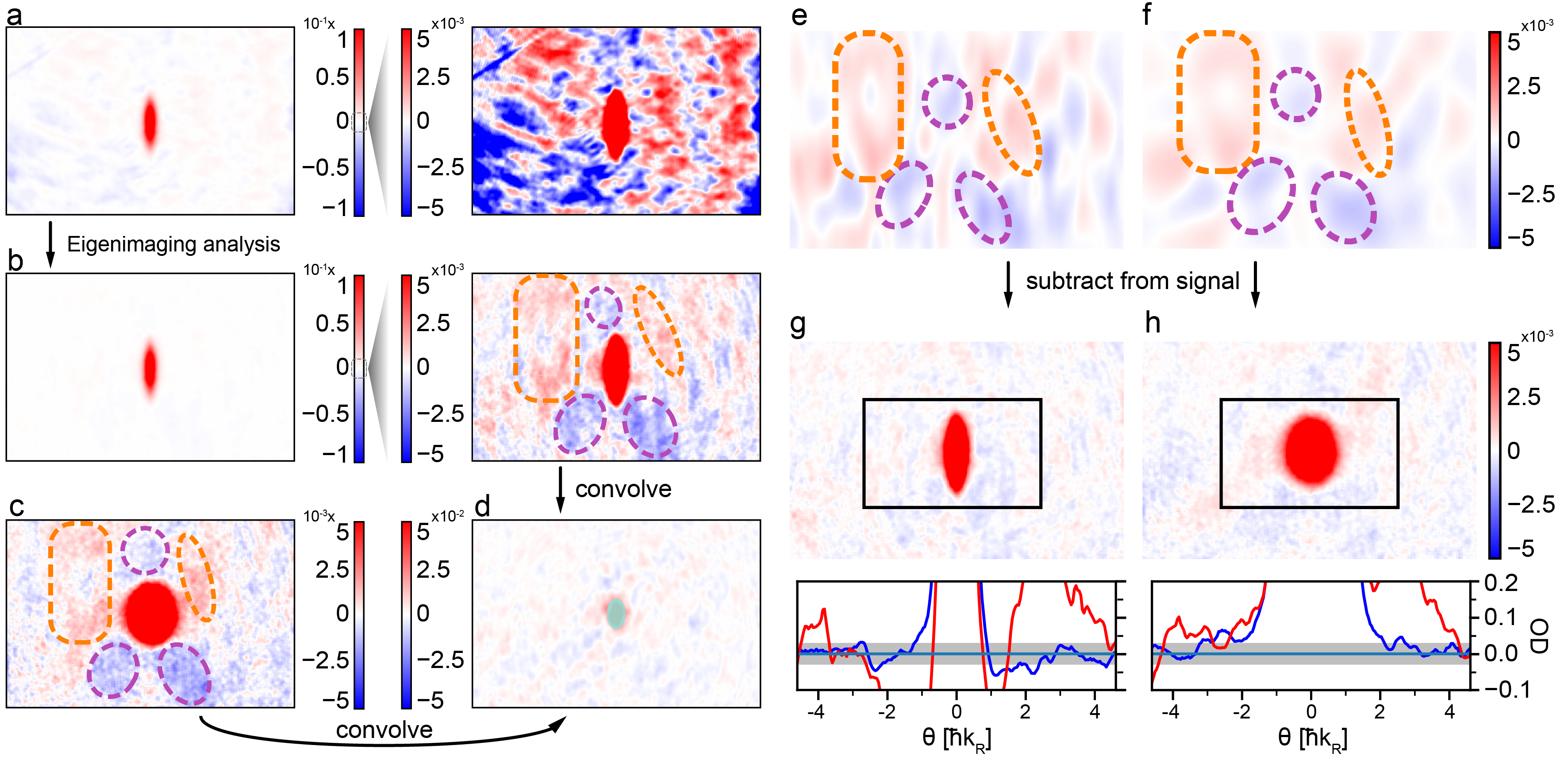}
     \caption{Two-step noise removal protocol for the high-resolution data. (a), Left: 2D profile before eigenimage analysis for the scar state at $t'=$0~ms. The SNR is on the order of $10^2$. Right: Same profile plotted under a saturated scale to showcase the background noise level.(b), Left: 2D profile after eigenimage analysis for the scar state at $t'=$0~ms. The noise level is reduced by a factor of approximately 4. Right: Same profile plotted under a saturated scale. The residual low-frequency spatial noise is likely due to imaging artifacts caused by light indirectly scattering off the atoms, to various optics, and then back into camera. Orange (purple) dashed lines indicate characteristic highly positive (negative) noise regions. (c), Averaged 2D profile after eigenimage analysis for the scar state at $t’=$3.3~ms. (d), Residual background of a point source $\mathcal{R}(x)$ realized by the 2D profile of a small BEC. The center BEC indicated by the green oval is masked out in convolution. (e--f), The correction mask for the scar state at (e), $t’=0$~ms and (f), 3.3~ms. Orange (purple) dashed circles show corresponding positive (negative) features in (b) and (c). (g--h), 2D profile after subtracting correction mask for the (g), $t’=0$~ms and (h), 3.3~ms images. The black box shows the region of interest with momenta cutoff at $\pm 4.5 k_R$. The integrated 1D profile are shown below. Red (blue) line represents the signal before (after) image processing. The gray band around zero shows the 1-$\sigma$ statistical uncertainty from residual noise after image processing.}
    \label{fig:Eigenimage}
\end{figure*}

\subsection{Initial state preparation}

We prepare 1D dipolar gases with different Lieb-Liniger parameters $\gamma$ by ramping the magnitude of the bias magnetic field along $\hat{z}$. We perform the ramp fast enough to avoid significant atom loss due to inelastic three-body collisions when sweeping near the Feshbach resonance.  However, it is sufficiently slow that the system follows the holonomy cycle without trap excitation.  We experimentally determine the optimal ramp time by measuring the breathing amplitude after the field ramp.  We aim to maintain this within 5-10\% of the equilibrium gas width, which results in a field ramp duration of 1~ms between 26.69~G to 26.775~G for the $\gamma = 26.3$ state. Note that what we call $\gamma$ here and in the main text is what we will begin to denote as $\gamma_T$ below.

To reach states along the attractive branch of the holonomy cycle, we spend as little time as possible crossing the pole to smoothly transition into sTG state. This constraint, combined with minimizing trap excitation, motivates a two-step ramp protocol: First we slowly ramp in 30~ms into the vicinity of the CIR pole near the Feshbach resonance, and then cross the pole and ramp to the final field in 1~ms. This minimizes heating at the resonance.  The resulting breathing amplitude is no more than 5\% of the equilibrium gas width.

\subsection{Magnetic field calibration}

We calibrate the magnitude of the magnetic bias field on a daily basis by performing radio frequency spectroscopy to drive transitions between Zeeman levels~\cite{Kao2021tpo}.  We determine the field magnitude by scanning the radio frequency and fitting the atom loss features to a Gaussian line shape. The typical uncertainty of the resulting field is less than $\pm$2~mG.

\subsection{Measurements of full width at half maximum}

The shape of the rapidity distributions are shown in comparison with theory in Fig.~\ref{fig:FWHM}(a--c). For the FWHM data, we use a bootstrap method to obtain a nominal value and error for each data point. In addition to the FWHM data for the sTG state in Fig.~1b, Figs.~\ref{fig:FWHM}(d--f) show the FWHM of the remaining three initial states at each quench time step for the 10$\times$ quench experiment. In accordance with the main text, simulation curves are shifted with respect to the experiment curves by the same amount as in Fig.~3 of the main text.

\subsection{Image processing}

We employ a two-step noise removal protocol to our absorption images to achieve an optical density (OD) accuracy at the $10^{-3}$ level. The absorption images are first processed by eigenimage analysis to remove shot-to-shot variation of high-frequency spatial interference patterns~\cite{Li2007roi}. Figures~\ref{fig:Eigenimage}(a,b)  compare 2D TOF profiles before and after eigenimage analysis. The processed images are then post-selected based on a threshold of $\pm 10$\% of the average atom number. Around 60 shots remain after post-selection for each time step in the data shown in Figs.~2 and~3, as well as in Figs.~\ref{fig:Parameters}(f--g),~\ref{fig:FWHM}, and~\ref{fig:2xquench}.  These are averaged to obtain a 2D OD image, such as that shown in Fig.~\ref{fig:Eigenimage}(c). 

We follow the same protocol for background removal as in Ref.~\cite{Li2023ram}. First we integrate over the $\hat{z}$ to obtain a 1D OD profile. Then we identify the background noise region by calculating where the average OD is within $\pm 1 \sigma$ of zero.  Here, $\sigma$ denotes the pixel-wise statistical standard deviation of the extracted OD due to photon shot noise and CCD dark counts. We remove the background systematic noise by subtracting a third-order polynomial fit to the noise. 

The high-resolution data in Fig.~4---from the average of hundreds of shots---requires a bit more care because we are interested in signal at high-$k$, where the SNR approaches unity.  To avoid overfitting, we use a different removal method for reducing residual background. This applies a pixel-wise correction to the averaged 2D profile. The residual is attributed to imaging artifacts  that consists of mainly low-frequency spatial noise.
This is likely caused by light from the atoms reflecting off the imaging optics in a way that is different from the desired transmitted scattering.  Such background is present in all atomic images and is not corrected by background subtraction or the eigenimage analysis. It does not seem to significantly depend on the size of the atomic clouds employed, allowing us to use the same removal procedure for all $t'$.  We can model this by a point spread function containing features of the background. The model is motivated and supported by empirical observations that features in Figs.~\ref{fig:Eigenimage}(a--c) move along with atoms, rather than remain static on the camera when we change the atom location by ${\sim}10$~$\mu$m. The 2D profile after eigenimage analysis can be written as:
\begin{equation}
\text{OD}_\text{meas}(x) = \text{OD}(x) + \text{OD}(x) * \mathcal{R}(x),
\end{equation}
where $*$ denotes convolution, $\text{OD}_\text{meas}(x)$ is the measured 2D profile after eigenimage analysis, OD$(x)$ is the true OD proportional to the atomic density, and $\mathcal{R}(x)$ represents the residual background scattered off a point atomic source. The background of a point source $\mathcal{R}(x)$ is estimated by first taking images of a small BEC with atom number around $0.6(1)\times 10^4$ and then masking out the BEC region, as shown in Fig.~\ref{fig:Eigenimage}(d). $\mathcal{R}(x)$ is typically on the $10^{-3}$ scale, comparable to the high-$k$ signal of interest.

We obtain $\text{OD}(x)$ from measured 2D profiles through the following calculation:
\begin{align}
    \text{OD}_\text{meas}(x) &- \text{OD}_\text{meas}(x) * \mathcal{R}(x) \nonumber \\ 
    &= \text{OD}(x) - \text{OD}(x) * \mathcal{R}(x) * \mathcal{R}(x).
\end{align}
where the second term $\text{OD}(x) * \mathcal{R}(x) * \mathcal{R}(x) \sim 10^{-6}$ is orders-of-magnitude smaller than the residual background and can be neglected. The correction mask $\text{OD}_\text{meas}(x)*\mathcal{R}(x)$ shares many common features with the 2D profile after eigenimage analysis, as plotted in Figs.~\ref{fig:Eigenimage}(e-f). We subtract the correction mask and obtain the true signal. Figures~\ref{fig:Eigenimage}(g--h) show the corrected signal $\text{OD}_\text{meas}(x) - \text{OD}_\text{meas}(x) * \mathcal{R}(x)$ as well as the comparison of integrated 1D profiles before and after image processing for the region of interest in the main text. The resulting profile after the two-step protocol achieves a background noise floor of $10^{-3}$.

\begin{figure*}[t!]
    \centering
    \includegraphics[width=1\textwidth]{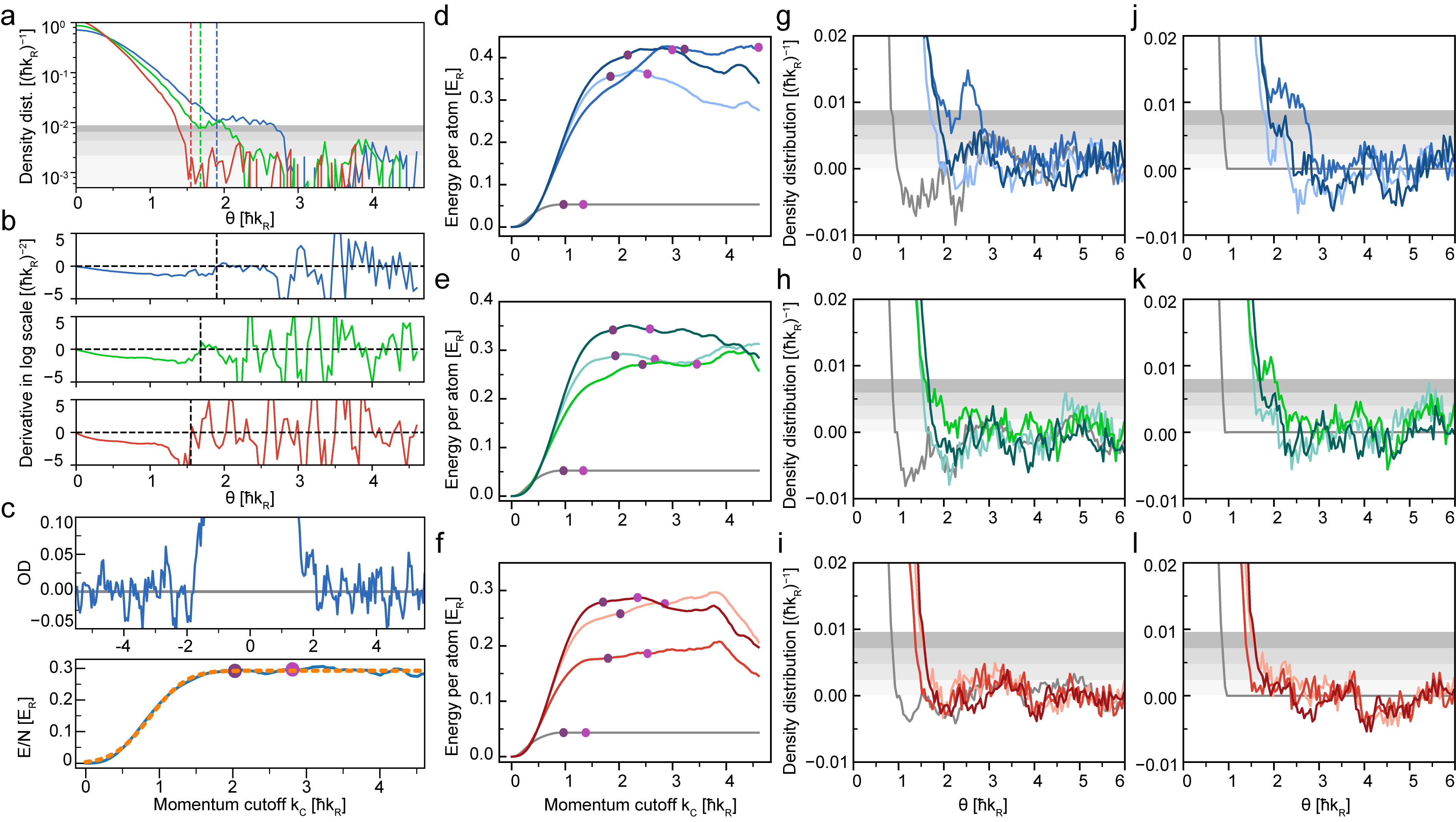}
    \caption{Momentum cutoff and energy integration.  (a--b) Determination of the cutoff point for momentum tail. (a), Momentum distribution at the maximal compression time for the scar state (blue), sTG state limit (green), and repulsive state (red). Momentum cutoff separating the body from the tail is shown as a vertical dashed line of the corresponding color. (b), Derivative of the momentum distribution on a log scale.  The cutoff is defined as the momentum at which the derivative first crosses zero. (c), Kinetic energy per atom for the scar state at $t’=$3.3~ms using an average of only 60 shots. Top: 1D profile of momentum distribution after image processing. Bottom: Energy per atom versus integration cutoff. Energy integration and the corresponding error function fit are shown in blue solid and orange dashed curves. Left cutoff (dark purple dot) and right cutoff (purple do) are marked at 3$\sigma_\text{fit}$ and 5$\sigma_\text{fit}$, resp. The plateau is flat and the resulting uncertainty from the remaining background is small. (d--f), Kinetic energy per atom is integrated up to different momenta $\theta$ for each of the traces.  The value used for the plots in the main text is determined by the average value between the left cutoff (dark purple dot) and right cutoff (purple dot).  (g--l), Low-noise-floor momentum measurements from hundreds of shots.  (g--i), Momentum distributions for the (g), scar; (h), sTG; and (i), repulsive states. As in Fig.~4 of the main text, the three color shades in each panel show the momentum distribution at $t' = 2.3$~ms (lightest), 3.3~ms (medium), and 4.3~ms (darkest shade). The gray line corresponds to the momentum distribution at $t' = 0$~ms. Gray bars represents noise floor from 1$\sigma$ to 4$\sigma$ ordered bottom to top. (j--l), Same as before, but with data normalized by the $t' = 0$~ms distribution. The subtraction begins where the gray trace flattens to zero and extends throughout the rest of the high-$k$ data. (The $t'=0$ data flattens to zero where we subtract it from itself.) Panels (j) and (l) are the same as Figs.~4(a,b) of the main text, but plotted on a linear scale (and including $t’=0$~ms data.) }
    \label{fig:EnergyTail}
\end{figure*} 

\begin{figure*}[t!]
    \centering
    \includegraphics[width=0.9\textwidth]{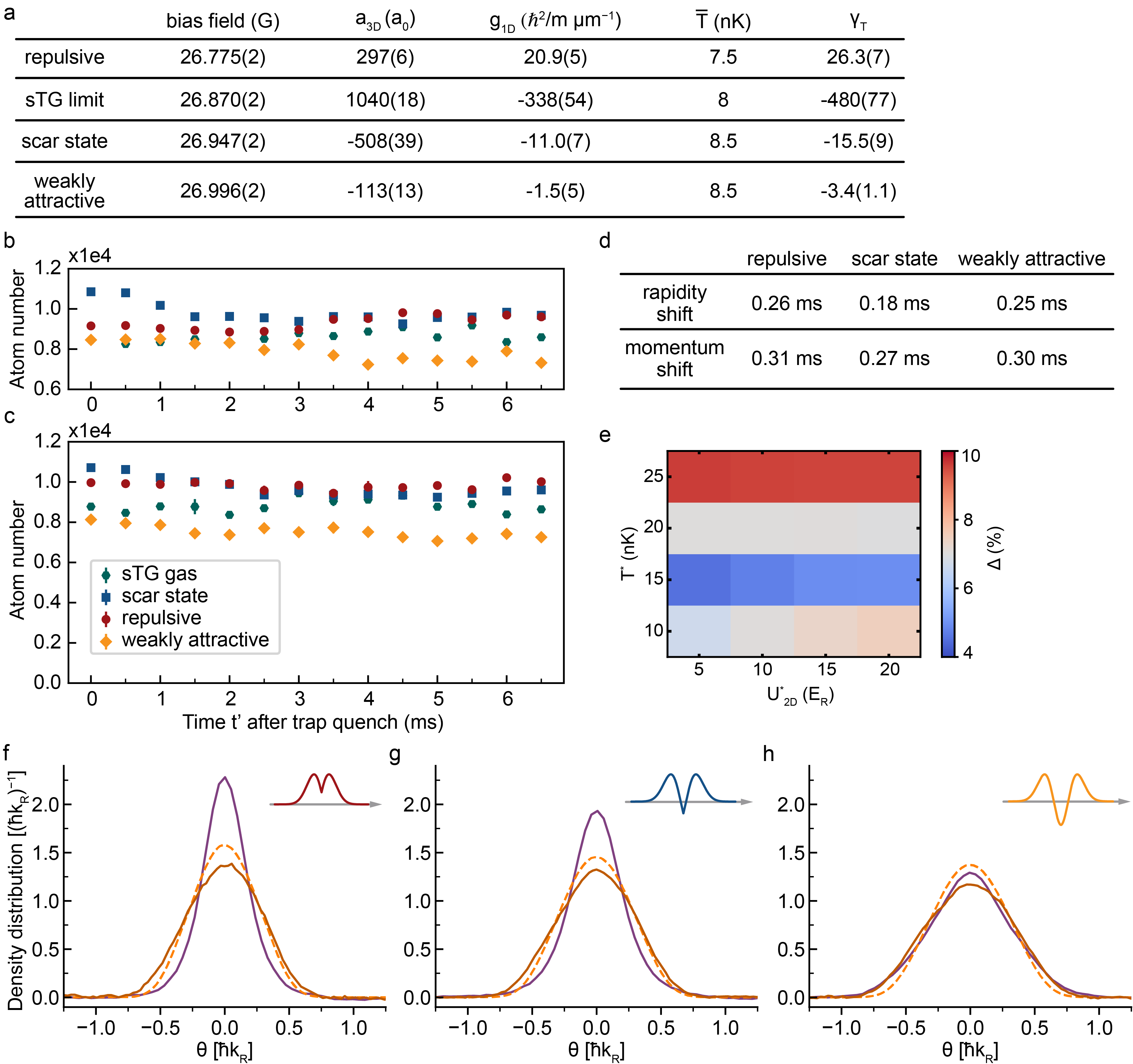}
    \caption{Experiment and simulation parameters. (a) Magnetic field and interaction parameters for each of the states studied in the main text. $a_{3D}$ and $g_{1D}$ uncertainties come from the magnetic bias field. See Sec.~\ref{sec:gamma_T} for the definition and error analysis for $\bar{T}$ and $\gamma_{T}$. (b--c) Total number of atoms in the optical lattice at each time $t’$ for the 10$\times$ quench for (b), momentum and (c), rapidity measurements, resp. (d) Time shifts applied to simulation curves in each panel of Fig.~3. See Sec.~\ref{sec:TGLimit} for a discussion of these shifts. (e) Theoretical error $\Delta$ as a function $U_{\rm 2D}^*$ and $T^*$ for the initial state ($t'=0$) of the sTG-state quench. We use the optimized value $U_{\rm 2D}^*=5E_{R}$ and $T^*=15$~nK to compute the initial states for all quenches in the main text. (f--h) Initial state at $t’=0$ for (f), repulsive; (g), scar; and (h), weakly attractive states. Experimental momentum (rapidity) distribution are shown as solid purple (orange) curves. Numerical rapidity distributions calculated using parameters in panel (e) are shown as dashed orange curves.}
    \label{fig:Parameters}
\end{figure*} 

\subsection{Partition of momentum distribution into low and high-$k$ regions}

Figures~\ref{fig:EnergyTail}(a) shows the momentum at which we notionally partition the momentum distributions into low and high-$k$ regions. This partition cutoff point distinguishes the low-$k$ data shown in Fig.~3 from the higher-resolution data that includes the high-$k$ tails in Fig.~4. We do this for each state by first taking the derivative of each distribution. This is shown in log scale in Fig.~\ref{fig:EnergyTail}(b). The high-$k$ region is defined as the point at which we observe the onset of the momentum shoulder or flat background. This point is determined by finding where the derivative crosses zero. 

\subsection{Kinetic energy per atom versus momentum cutoff}

We sum the energy from the distribution center to a finite momentum cutoff $k_c$. The average energy per atom is then given by
\begin{equation}
E_{k_c}=\sum_{k \in\left[-k_c, k_c\right]} \frac{k^2}{2m} f\left(k\right) \Delta k.
\label{eq:energy_sum}
\end{equation}
The energy per atom increases as a function of the integration cutoff $k_c$ before eventually reaching a plateau. We then fit the integrated energy versus $k_c$ to an error function $E_{k_c} = a \cdot [\text{Erf}(b \cdot (k_c-c)) + 1] / 2$, where Erf($z$) is the Gaussian error function $\text{Erf}(z)=2/\sqrt{\pi} \int_0^z e^{-t^2} \mathrm{~d} t$; see Fig.~\ref{fig:EnergyTail}(a). The fit parameter $a$ guesses the true energy while $1/(\sqrt{2}b)$ and $c$ represents the standard deviation $\sigma_\text{fit}$ and mean $\mu$ of the Gaussian fit. As in Ref.~\cite{Malvania2021ghi}, the energy per atom is taken to be the average of $E_{k_c}$, where $k_c \in [\mu+3\sigma_\text{fit}, \mu+5\sigma_\text{fit}]$. This interval is chosen to ensure that the energy has sufficiently reached the plateau but also not extended too far into the regime dominated by background noise. 

There are two sources of error in $E_c$. The first is from the remaining background which appears as a deviation from a flat $E_{k_c}$ at high momentum. This error can be estimated by the peak-to-peak difference in energy from the $k \in [\mu+3\sigma_\text{fit}, \mu+5\sigma_\text{fit}]$ region and is a small uncertainty in this 60-shot case. The second source is related to the uncertainty of background shape in the signal region. As for atom number and FWHM measurements, we estimate this error by bootstrapping 1000 sets of samples from the original data set. 

For the high-resolution data, we average over all shots to obtain a 1D momentum distribution and report the average $E_{k}$ value between $k\in [\mu+3\sigma_\text{fit}, \mu+5\sigma_\text{fit}]$.  Figures~\ref{fig:EnergyTail}(d--f) show the average energy per atom plotted against momentum cutoff $k_c$. Unlike the 60-shot data sets, ripples from the remaining background noise at higher $k$ causes more fluctuations in $E_{k_c}$. Consequently, the peak-to-peak difference between $[\mu+3\sigma_\text{fit}, \mu+5\sigma_\text{fit}]$ region is no longer negligible.  We assign an error $\Delta E_{k_c}$ to this by replacing $f(k)$ in  Eq.~\eqref{eq:energy_sum} with the noise floor and choosing $k_c = \mu+3\sigma_\text{fit}$ as integration limit. The final uncertainty reported in the main text is the quadrature sum of both error sources.

\subsection{Background noise-floor subtraction}

On an absolute scale, the noise floor for the data in Fig.~4 (averaged hundreds of times) is $\sim$3$\times$-lower than the 60-shot data set in Fig.~3.  However, there remains a systematic background.  Therefore, we choose to present these data relative to the distribution at $t'=0$~ms.  Thus, the $t'=0$~ms data serves as a background reference for all other quench times, for each state. We identify the noise cutoff in momentum space at which the $t'=0$~ms distribution first crosses zero and subtract this noise from the momentum distributions taken at $t'>0$. Finally, we symmetrize the distributions with respect to the center to obtain the momentum distributions in Fig.~3 of the main text. The reported $\sigma$ of the noise floor for each state is determined by calculating the standard error from the $t'=0$~ms signal between 4~$k_R$ to 6~$k_R$. Figures~\ref{fig:EnergyTail}(g--l) show the 1D momentum distributions for each state before and after subtracting the 0-ms background. 

\begin{figure*}[t!]
    \includegraphics[width=1\textwidth]{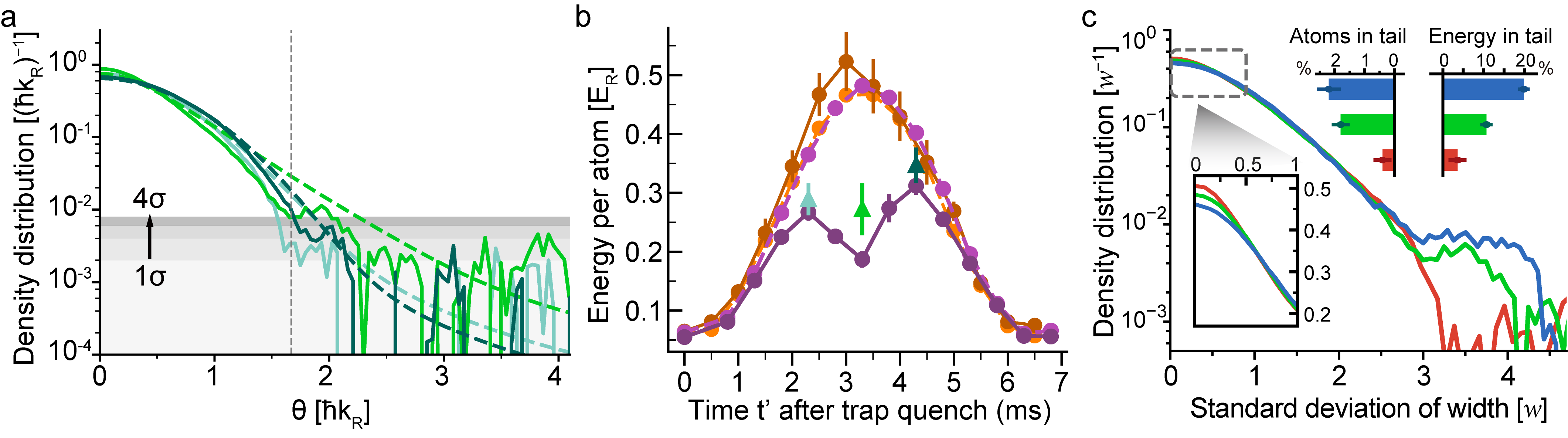}
    \caption{High-resolution momentum distribution measurements for sTG gas.  (a), Momentum density distribution (average of 600 shots) at $t’=2.3$~ms (greenish blue), 3.3~ms (green), and 4.3~ms (dark green). Dashed curves correspond to the simulation results at the same $t’$. The gray horizontal color bands mark the statistical uncertainty from the mean background noise in steps of one-to-four standard error ($\sigma$). The gray vertical dashed lines indicate the cutoff in $\theta$ between the low and high-$k$ region. (b), Time evolution of kinetic (purple) and total (orange) energy. Data and simulation results are shown in solid and dashed curves, respectively. Triangular markers indicate the integrated kinetic energy of the curve with corresponding color shade in panel (a). (c), Momentum distribution for $t’=$3.3~ms normalized by the standard deviation of each curve's width $w$. The scar (blue) and repulsive (red) states are added for comparison; these are identical to those in Fig.~4(c). Lower inset: Zoomed-in view of the low-momentum peaks. Upper inset: Percentages of atom number and kinetic energy in the high-$k$ region beyond the partition indicated by the vertical line in panel (a) for the sTG state and in Figs.~4(a,b) for the repulsive and scar states, resp. }
    \label{fig:sTGHighRes}
\end{figure*}

\subsection{High-resolution data for the Super-Tonks-Girardeau state}

In addition to low-noise floor measurements for repulsive and scar state, shown in Figs.~4(a,b), we present high-resolution data for the sTG gas in Fig.~\ref{fig:sTGHighRes}. Over 600 shots are averaged to achieve a noise floor of ${\sim}2\times 10^{-3}$. The momentum distribution can be calculated by using hardcore bosons at the sTG limit. See Sec.~\ref{sec:TGLimit} for details of the calculation. This enables a direct comparison of experiment and simulation results. From Fig.~\ref{fig:sTGHighRes}(a), we see that experiment matches the simulation trend. First, the momentum distribution at maximum compression time $t'=$3.3~ms is taller than those at 2.3~ms and 4.3~ms. Second, the tails of the distribution at $t'=$3.3 ms extends further into high-$k$ region as a result of bosonization at high atomic density~\cite{Wilson2020ood}. We observe a short shoulder up to $\theta \approx 2\hbar k_R$. 

We plot the time evolution of the energy per atom in Fig.~\ref{fig:sTGHighRes}(b). Total energy via rapidity matches simulation throughout the first cycle.  There is an energy gap near the maximum compression point for kinetic energy. Part of this is due to atoms in the high-$k$ tail, which we uncover using these low-noise floor measurements, which are plotted as triangles. As discussed in the main text, the remaining missing energy may be attributed to the long-range DDI, which may account all of the 0.2$E_R$ gap. Another possibility is the long tail under noise floor due to bosonization. This is shown to be lower and shorter than the scar state in Fig.~\ref{fig:sTGHighRes}(c). 

\begin{figure*}[t!]
    \includegraphics[width=1\textwidth]{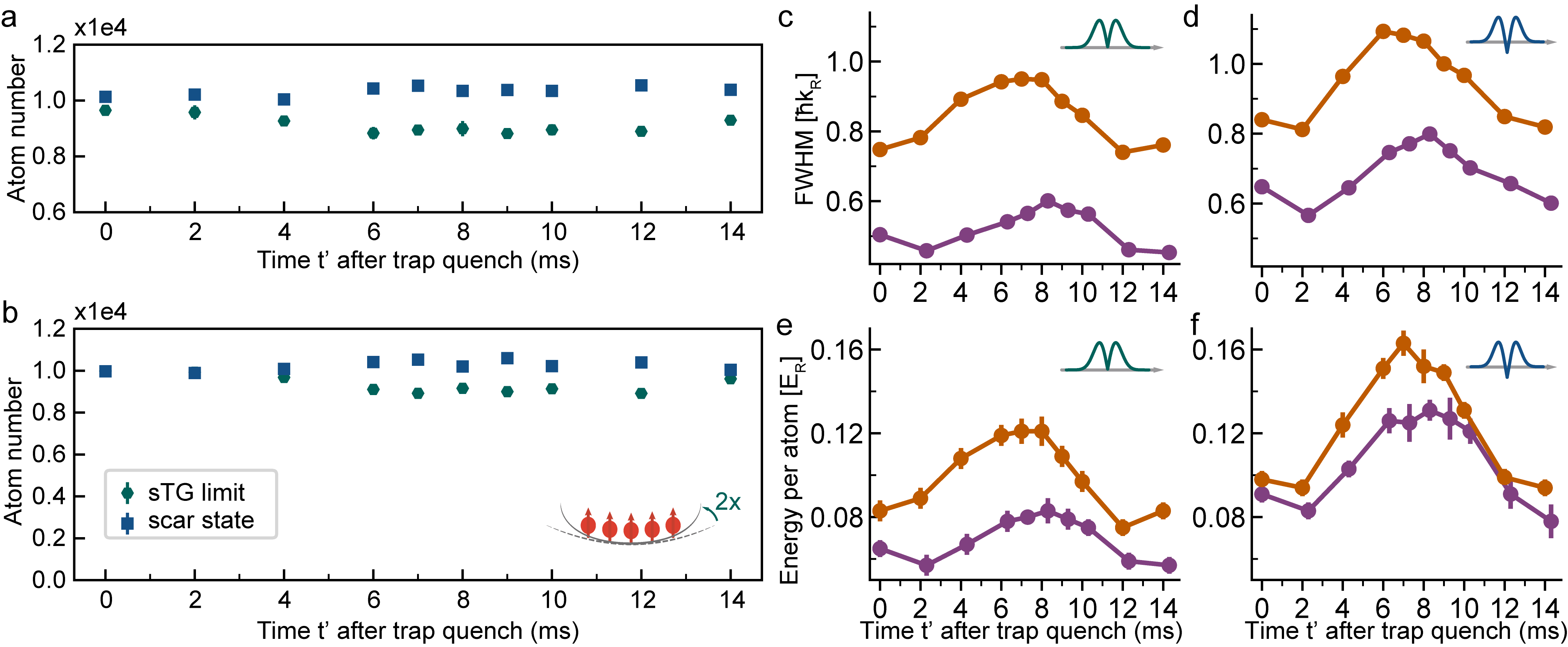}
    \caption{2$\times$ quench experiment. (a--b) Total number of atoms in the optical lattice at each time $t’$ after the 2$\times$ quench: (b), momentum and (c), rapidity measurements. Blue squares (green hexagons) are for the scar (sTG) state. (c--f) Experimental results for the 2$\times$ trap quench experiments. (c--d), Time evolution of FWHM over the first cycle after the quench for (c), sTG and (d), scar states. Maximal compression occurs between 6-8~ms. Purple (orange) curve shows results from momentum (rapidity) measurements. Data points are the average of 60 shots. (e--f), After-quench dynamics of the energy per atom over the first cycle. Purple (orange) curve represents kinetic (total) energy per atom.  There is no dip near maximal compression for momentum distribution.}
    \label{fig:2xquench}
\end{figure*} 

\subsection{Scattering length determination}\label{sec:FR}

Fitting parameters extracted from molecular binding energy and atom loss spectroscopy measurements in Ref.~\cite{Kao2021tpo} are used to model the four Feshbach resonances involved in calculating the $s$-wave scattering length $a_{3D}$. The values of magnetic field we used and their corresponding $a_{3D}$ are listed in Fig.~\ref{fig:Parameters}(a). 

\subsection{Atom number measurements}

Figures~\ref{fig:Parameters}(b,c) show the total atom number in the optical lattice versus $t'$ of all four initial states during the first oscillation cycle. Both the data points and error bars are obtained through bootstrapping, where instead of fitting the background from the averaged 1D OD profile, we randomly sample from the set of all shots to select 60 for fitting and background removal. Atom number is calculated by integrating the resulting OD profile. We then repeat the sampling 1000 times to obtain the atom number distribution.

\subsection{2$\times$ quench experiment}

We perform gentler trap quench to demonstrate that the phantom energy phenomenon is a nonlinear effect. We follow the exact same experimental sequence as in the 10$\times$-quench experiment up until initial state preparation. But now we increase the trap potential by only 2$\times$, jumping the longitudinal trap frequency $\omega_{\parallel}$ from $2\pi\times$27.8(5)~Hz to $2\pi\times$39.8(3)~Hz. As before, we measure momentum and rapidity distribution over the first cycle, and the same lattice deloading and TOF expansion sequence is used.

We summarize these results in Fig.~\ref{fig:2xquench}. The atom loss over the first cycle is less than $10\%$ for both sTG gas and scar states, as shown in panels (a,b). After-quench dynamics for both FWHM and energy per atom are plotted in panels (c--f). While the rapidity measurement results (orange) follow the same sinusoidal trend as in 10$\times$ quench experiments, momentum measurement results (purple) behave very differently. For the sTG gas, the 2$\times$-quench does not induce a dip in FHWM or kinetic energy near maximum compression.  We also find that even for scar state, the 2$\times$-quench is gentle enough to not introduce a dip in FHWM or kinetic energy. This indicates that the phantom energy phenomenon is a nonlinear effect:  It  emerges only upon violent compression.

\section{Modeling}

\subsection{Hamiltonian}

The experiment is modeled as a 2D array of independent 1D gases, which we refer to as ``tubes'' in what follows. (We neglect the intertube DDI but estimate its effects below.) Each tube is described by the Lieb-Liniger Hamiltonian~\cite{Lieb1963eao_2} extended by the addition of a longitudinal harmonic confining potential, $U_{\rm H}$, and an intratube DDI, $U^{\rm 1D}_{\rm DDI}$,
\begin{equation}\label{eq:H_LL}
H=\sum_{i=1}^{N}\left[-\frac{\hbar^2}{2m}\frac{\partial^2}{\partial x_i^2}+U_{\rm H}(x_i)\right] +\sum_{1\leq i<j \leq N}\left[g_{\rm 1D}^{\rm vdW}\delta(x_i-x_j)+U^{\rm 1D}_{\rm DDI}(\theta_\text{B}, x_i-x_j)\right].
\end{equation}
$N$ is the number of atoms within a tube, $m$ is the mass of a $^{162}$Dy atom, and $g_{\rm 1D}^{\rm vdW}=-2\hbar^2/(ma_{\rm 1D})$ (both repulsive and attractive in the experiment) is the effective 1D contact interaction due to the van der Waals interaction. $g_{\rm 1D}^{\rm vdW}$ depends on the 3D s-wave scattering length $a_{\rm 3D}(B)$ (set by the magnetic field $B$) and on the depth of the 2D optical lattice $U_{\rm 2D}$, through the 1D scattering length $a_{\rm 1D}=-a_{\perp}[a_{\perp}/a_{\rm 3D}(B)-C]/2$. $a_{\perp}=\sqrt{2\hbar/(m\omega_{\perp})}$ is the transverse confinement, $\omega_{\perp}=\sqrt{2U_{\rm 2D}k_R^2/m}$, and $C=-\zeta(1/2)\simeq1.4603$~\cite{Olshanii1998asi}. The longitudinal harmonic confinement potential $U_{\rm H}(x)=\frac{1}{2}m\omega_x^2 x^2$, where $\omega_x$ is the longitudinal trapping frequency. For the trap quench experiments, $\omega_x$ changes from its initial value (at the end of the state preparation) $\omega_{x}^i$ to $\omega_{x}^f$ at $t'=0$.

The intratube DDI in the single-mode approximation can be written as~\cite{Deuretzbacher2010gpo, Deuretzbacher2013egp,Tang2018tni,Kao2021tpo,DePalo2021pad},
\begin{equation}
U^{\rm 1D}_{\rm DDI}(\theta_\text{B}, x)=\frac{\mu_0\mu^2}{4\pi}\frac{1-3\cos^2\theta_\text{B}}{\sqrt{2}a_{\perp}^3}\bigg[V^{\rm 1D}_{\rm DDI}(u)-\frac{8}{3}\delta(u)\bigg]\,,
\end{equation}
where $V^{\rm 1D}_{\rm DDI}(u)=-2|u|+\sqrt{2\pi}(1+u^2)e^{u^2/2}{\rm erfc}(|u|/\sqrt{2})$, $u=\sqrt{2}x/a_{\perp}$, and ${\rm erfc}(x)$ is the complementary error function. $\mu=9.93\mu_\text{B}$ is the dipole moment of $^{162}$Dy. As in Ref.~\cite{Li2023ram}, we account for the leading-order effect of $U^{\rm 1D}_{\rm DDI}(\theta_\text{B}, x)$ by treating it as a contact interaction term,
\begin{equation}\label{UthetaB}
\tilde U(\theta_\text{B}, x)=\frac{\mu_0\mu^2}{4\pi}\frac{1-3\cos^2\theta_\text{B}}{2a_{\perp}^2}\left[A-\frac{8}{3}\right]\delta(x)=g_{\rm 1D}^{\rm DDI}\delta(x)\,,
\end{equation} 
where $A=\int_{-\infty}^{\infty}V^{\rm 1D}_{\rm DDI}(u)du=4$. After this simplification, the Hamiltonian~\eqref{eq:H_LL} can be written as
\begin{equation}\label{eq:H_LL2}
\tilde H=\sum_{i=1}^{N}\bigg[-\frac{\hbar^2}{2m}\frac{\partial^2}{\partial x_i^2}+U_{\rm H}(x_i)\bigg]+\sum_{1\leq i<j \leq N}g_{\rm 1D}\delta(x_i-x_j)\,,
\end{equation}
where 
\begin{equation}\label{eq:g1d}
g_{\rm 1D}=g_{\rm 1D}^{\rm vdW}+g_{\rm 1D}^{\rm DDI}.
\end{equation}
For repulsive interactions ($g_{\rm 1D}>0$) in the absence of the trapping potential $U_{\rm H}$, the Hamiltonian~\eqref{eq:H_LL2} can be solved exactly using the Bethe ansatz~\cite{Lieb1963eao_2,Yang1969toa}. For attractive interactions ($g_{\rm 1D}<0$), we focus on the highly excited `super-Tonks-Girardeau' (sTG) gas states \cite{Astrakharchik2004qbg, Astrakharchik2005btt,Batchelor2005eft, Haller2009roa,Chen2010tfa,Kao2021tpo}. They are obtained from real solutions of the Bethe ansatz equations for $g_{\rm 1D}<0$~\cite{Chen2010tfa}, and are realized in our experimental setup at $\theta_{\rm B}=90^\circ$~\cite{Kao2021tpo} using a topological pumping protocol~\cite{Yonezawa2013qhi,Kasumie2016aeo,Kao2021tpo}.

\subsection{Initial state preparation}\label{sec:statprep}

To describe the initial state right before our trap quenches, we need to find the atom number $N_l$ and temperature $T_l$ of each 1D tube ``$l$" at position ($y_l$, $z_l$) in the 2D optical lattice. $N_l$ and $T_l$ are computed using the same state preparation modeling as in Ref.~\cite{Li2023ram}, which we summarize in this section for completeness.

In the experiment, a 3D BEC is loaded into a 2D optical lattice ($U_{\rm 2D}$), which is ramped up to create a 2D array of 1D tubes. We make the following assumptions: (1) At $U^*_{\rm 2D}$ (which sets $g^*_{\rm 1D}$), the entire 3D system decouples into individual 1D tubes with $N_l$ atoms. (2) At ``decoupling,'' all tubes are in thermal equilibrium at the same temperature $T^*$. (3) As the loading proceeds beyond $U^*_{\rm 2D}$, the process is adiabatic, i.e., the entropy for each tube remains constant. We determine the atom number $N_l$ and entropy $S_l$ of each tube $l$ by using the exact solution of the homogeneous Lieb-Liniger model at finite temperature plus the local density approximation (LDA). This relies on the first two assumptions and having in hand the experimental total atom number $N_{\rm tot}$ and the values of the ODT frequencies $\omega_x$, $\omega_y$, and $\omega_z$ during the lattice loading.

Using the third assumption and the experimental parameters at the end of the loading process, we obtain the temperature $T_l$ (and the corresponding chemical potential $\mu_l$) of each tube. As a simplification, we group the 1D tubes with the same $N_l=N$ (rounded to the closest integer) and assume that they have the same entropy in our calculation, using the average entropy $\bar S(N) = \overline{ S_l(N_l=N) }$. We then search for $T(N)$ (in a grid of temperatures that change in steps of 0.5~nK) that produces the appropriate $\bar S(N)$. This relies on the value of $\omega_x^i$ and $g_{\rm 1D}$ at the end of the state preparation. Note that for attractive interactions, the ``temperature'' and ``entropy'' we find are well defined because we consider only sTG states.

The free parameters for our modeling are $U^*_{\rm 2D}$ and $T^*$. We optimize their values by minimizing the quadrature sum  $\Delta$ of the differences between the experimental measurements for the rapidity and momentum distributions in the initial state at $g_{\rm 1D}=-338\hbar^2/m$~$\mu$m$^{-1}$ (sTG state) and the corresponding theoretical calculations (in the TG limit, explained in what follows):
\begin{equation}
\Delta=\sqrt{\Delta^2_{\rm rapidity}+\Delta^2_{\rm momentum}}\,,
\end{equation}
where 
\begin{equation}
\Delta_{\alpha}=\frac{\sum |f_{\alpha}^{\rm exp.}(k)-f_{\alpha}^{\rm theo.}(k)|\delta k}{\sum |f_{\alpha}^{\rm exp.}(k)|\delta k+\sum |f_{\alpha}^{\rm theo.}(k)|\delta k}\,,
\end{equation}
and $\alpha$ denotes either rapidity or momentum. Note that momentum focusing is used for the momentum measurement of the initial state, so we do not need to consider the effect of TOF in this comparison. We call $\Delta$ the theoretical error. In Fig.~\ref{fig:Parameters}(e), we show $\Delta$ as a function of $U^*_{\rm 2D}$ and $T^*$, using the same grid as in Ref.~\cite{Li2023ram}; i.e., 5$E_R$ steps in decoupling depths and 5-nK steps in temperature. We choose as optimal values $U^*_{\rm 2D}=5E_R$ and $T^*=15$ nK.  We use the same set of $U^*_{\rm 2D}$ and $T^*$ optimized for the sTG quench for all other quenches, since the experimental state preparation follows the same protocol up to the point at which $U_{\rm 2D}$ reaches its final value and the total number of atoms are similar. Figures~\ref{fig:Parameters}(f--h) provide a direct comparison at $t'=$0~ms between simulation results (dashed) using these parameters and experimental data (solid).  

\subsection{Definition and values of $\gamma$}\label{sec:gamma_T}

Due to the presence of the confining potential in the experiments, $\gamma(x)=mg_{\rm 1D}/[\hbar^2n_{\rm 1D}(x)]$ depends on the local 1D density $n_{\rm 1D}(x)$. In such systems, it is common to compute the weighted average $\bar \gamma = \int dx n_{\rm 1D}(x) \gamma(x) / [\int dx n_{\rm 1D}(x)] = {\int dx g_{\rm 1D}}/N$. This quantity is well defined at zero temperature, for which $\int dx=x_0$, where $x_0$ is the size of the trapped 1D gas. However, at finite temperature, the particle density exhibits long tails and so $x_0$ is not well defined~\cite{Xu2015uso}. 

As in Ref.~\cite{Li2023ram}, we instead compute a $\gamma_T$ that is based on the ratio of the kinetic and interaction energy. For notational simplicity, this is what we call $\gamma$ in the main text. It is computed as follows. For a given set of experimental parameters, we calculate the ratio between the kinetic ($E_K=\sum_l E^l_K$) and interaction energy ($E_I=\sum_l E^l_I$), as obtained from our modeling. $\gamma_T$ is the value of the Lieb-Liniger parameter of a homogeneous system at finite temperature that has exactly that ratio. The homogeneous system is selected to have the same $g_{\rm 1D}$ as the trapped one, and a temperature that is the weighted average temperature of the array of 1D gases, $\bar T=\sum_l N_l T_l/N_\text{tot}$, where $N_\text{tot}=\sum_l N_l$. Since $E_K/E_I$ is a monotonic function with $\gamma$, one can always find the particle density $n^{\bar T}_{\rm 1D}$ for which $E_K/E_I$ matches the result obtained for the modeling of the experimental results. With it, we compute $\gamma_T=mg_{\rm 1D}/(\hbar^2n^{\bar T}_{\rm 1D})$. The values of $\gamma_T$ for the quenches that are considered in the main text are shown in Fig.~\ref{fig:Parameters}(a). They are highlighted in Fig.~1(b), where we plot the total energy per particle $(E_K+E_I)/N_\text{tot}$ as a function of $|\gamma_T|$. The total energy per particle in Fig.~1(b) exhibits some small jumps because of the 0.5~nK temperature discretization.

There are two sources of uncertainty in $\gamma_T$ that we consider. The first one comes from $g_{1D}$ via magnetic field uncertainty, as listed in Fig.~\ref{fig:Parameters}(a). The second one is the fluctuation of atom number from shot to shot, which enters through $n^{\bar T}_{\rm 1D}$. We measure an atom number standard deviation of $2\%$ for the several-hundred-shot data.  For simplicity, we both assume that this does not affect the volume of the ensemble and that the volume does not significantly change on its own. Thus, we linearly propagate the atom number uncertainty in quadrature sum with that from $g_{1D}$ to find the final error in $\gamma_T$, as listed in Fig.~\ref{fig:Parameters}(a).

\section{Numerical methods}

\subsection{Generalized hydrodynamics}\label{sec:GHD}

\begin{figure}[!t]
\includegraphics[width=0.45\columnwidth]{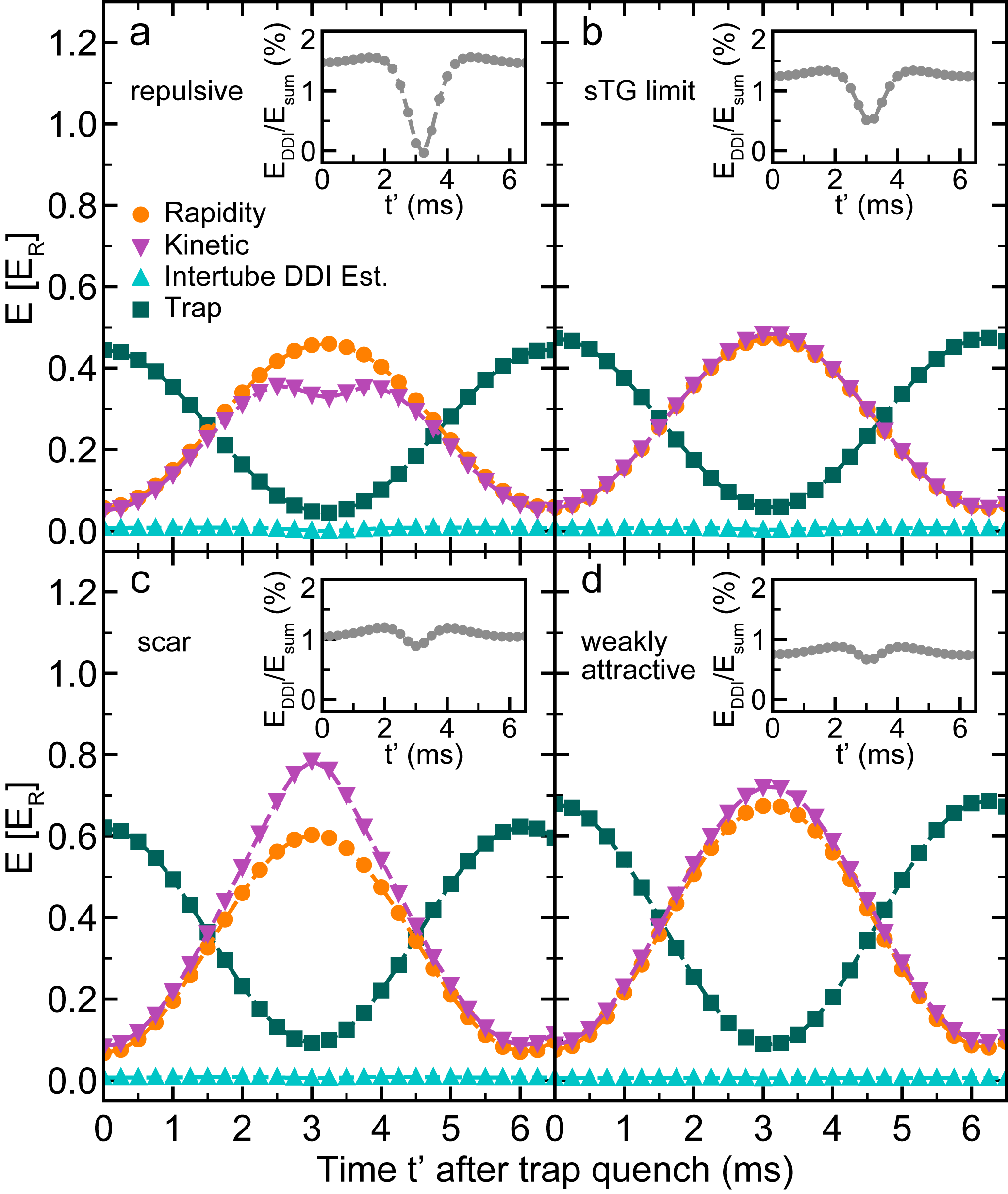}
\caption{Energies per particle from GHD simulations for the (a), repulsive; (b) sTG limit; (c), scar; and (d), weakly attractive states.  This is for a 10$\times$-trap quench. We show the total (rapidity) energy from Eq.~\eqref{eq:Erap}, kinetic energy from Eq.~(\ref{eq:Ek}), and the estimated intertube DDI and trap energy calculated from the GHD density distribution using Eq.~\eqref{eq:density}. Insets of each panel plot the ratio of the energy from the intertube DDI, $E_{\rm DDI}$, to the entire system, $E_{\rm sum}=E+E_{\rm trap}+E_{\rm DDI}$.}
\label{fig:DDI_estimate}
\end{figure}

The quench dynamics for a single tube $l$ can be simulated by solving the following generalized hydrodynamics (GHD) equations \cite{CastroAlvaredo2016ehi,Bertini2016tio,Doyon2017ano},
\begin{equation}\label{eq:GHD}
\partial_t \rho_l+\partial_x (v^\text{eff} \rho_l)= [\partial_x U^f_{\rm H}(x)/m] \,  \partial_{\theta} \rho_l\,,
\end{equation}
where $\rho_l(x,\theta,t)$ is the quasiparticle density with rapidity $\theta$ at position $x$ and time $t$. $v^{\rm eff}$ is the effective velocity for the quasiparticle $\theta$, which can be calculated by solving the following integral equation (at a given $x$ and $t$),
\begin{equation}
\label{eq:veff}
v^{\rm eff}(\theta) = \theta/m+ \int d\alpha\, \varphi(\theta-\alpha)\,\rho(\alpha)[v^\text{eff}(\alpha)-v^\text{eff}(\theta)].
\end{equation}
For the Lieb-Liniger model with $g_{\rm 1D}>0$,
\begin{equation}\label{eq:phi} 
\varphi(\theta-\alpha)=\frac{2mg_{\rm 1D}/\hbar}{(mg_{\rm 1D}/\hbar)^2+(\theta-\alpha)^2}.
\end{equation}
For the attractive case, $g_{\rm 1D}<0$, we assume that the dynamics after the quench involves only sTG states and the GHD equations remain the same but with a negative $g_{\rm 1D}$ in Eq.~\eqref{eq:phi}. 

Knowing $N_l$, $T_l$ for each tube, and the initial trapping frequency $\omega_x^i$, we calculate $\rho_l(x,\theta,t=0)$ by solving the thermodynamic Bethe ansatz equations \cite{Yang1969toa} within the LDA. Instead of directly time-evolving $\rho_l$ by means of Eq.~\eqref{eq:GHD}, we use the numerically more efficient molecular dynamics solver for GHD (the ``flea gas'' algorithm) introduced in Ref.~\cite{Doyon2018sga}. All GHD results shown in the main text were obtained using a time step $dt=5\times10^{-4}$~ms and an average over at least $10^6$ samples.

From $\rho_l(x,\theta,t)$, we calculate the averaged rapidity distributions,
\begin{equation}
    f(\theta,t)=\frac{1}{N_{tot}}\sum_l\int dx \rho_l(x,\theta,t)\,,
\end{equation}
the kinetic energy per particle,
\begin{equation}\label{eq:Ek}
    E_K=\frac{1}{N_{tot}}\sum_l\int dx d\theta \rho_l(x,\theta,t)\left[v^\text{eff}_l(x,\theta,t)-\frac{\theta}{2m}\right]\theta\,,
\end{equation}
and the total (rapidity) energy per particle,
\begin{equation}\label{eq:Erap}
    E=\frac{1}{N_{tot}}\sum_l\int dx d\theta \rho_l(x,\theta,t)\frac{\theta^2}{2m}\,.
\end{equation}
These are the observables for which our theoretical results are compared to the experimental measurements. We also compute the density distribution from GHD,
\begin{equation}\label{eq:density}
    n_l(x,t)=\int d\theta \rho_l(x,\theta,t)\,,
\end{equation}
which allows us to compute the trap energy, and estimate the intertube DDI energy, after the quench. In Fig.~\ref{fig:DDI_estimate}, we show the GHD simulations of rapidity, kinetic, intertube DDI, and trap energies for 10$\times$-quench experiments. For all cases we consider, the intertube DDI is $<2\%$ compared to the total energy scale of the system.

\begin{figure*}[!t]
\includegraphics[width=0.9\textwidth]{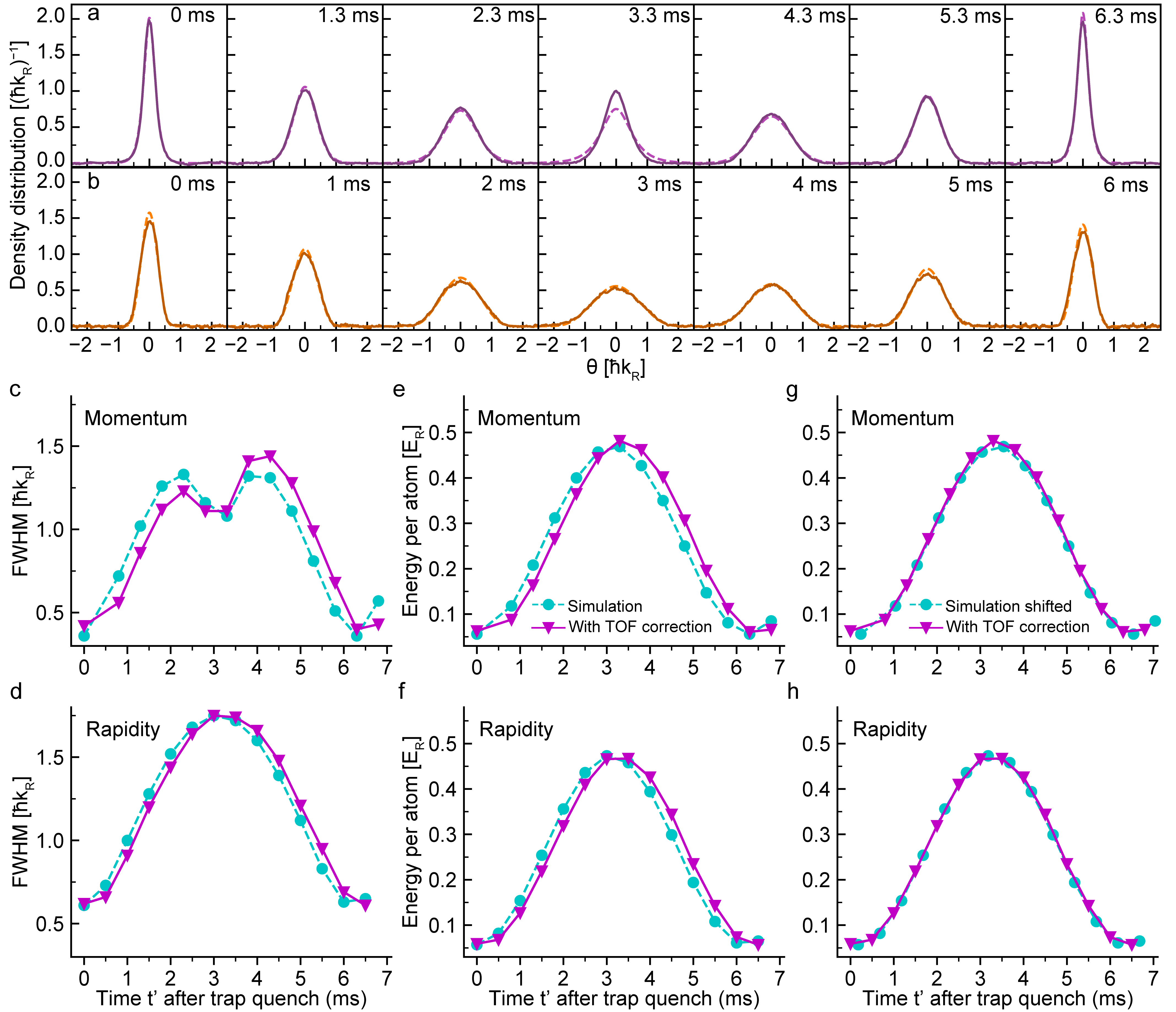}
\caption{Time-of-flight effects in the TG limit. Density distributions for (a), momentum and (b), rapidity at different $t'$ after the trap quench. (c,d) Direct comparison of the FWHM of the (c), momentum and (d), rapidity distributions with and without the TOF correction as functions of the evolution time in the trap after the quench. The TOF-corrected TG simulation is the dashed blue curve, while the sTG-state experimental data is solid purple. (e,f), Kinetic and total energies computed from the corresponding momentum and rapidity distributions, respectively. The TOF corrected simulation is in purple, while the bare simulation is in blue.  (g,h), We account for TOF effects by shifting the simulated kinetic and total energy curves. To do so, we follow the same procedure used in Fig.~3 in the main text. The kinetic energies are shifted by 0.24~ms and the rapidity energies by 0.15~ms.  The shifted simulation is in blue. The shift times are listed in Fig.~\ref{fig:Parameters}.}
\label{fig:TG_TOF}
\end{figure*}

\subsection{Tonks-Girardeau limit}\label{sec:TGLimit}

The equilibrium properties and the quench dynamics of Hamiltonian~\eqref{eq:H_LL2} can be exactly solved in the Tonks-Girardeau limit ($\gamma\to\infty$), which we use to benchmark  the experimental measurements of the sTG state in the main text. We use the following lattice hardcore boson Hamiltonian to carry out our calculations in that limit
\begin{equation}\label{eq:HCB}
    \hat H_\text{HCB}=-J\sum_{j=1}^{L-1}(\hat b^\dagger_{j+1}\hat b_j + {\rm H.c.})+ \sum_{j=1}^{L}U_{\rm H}(x_j)\hat b^\dagger_j\hat b_j\,,
\end{equation}
where $J$ is the hopping amplitude and $L$ is the total number of lattice sites. $\hat b^\dagger_j$ ($\hat b_j$) creates (annihilates) a hardcore boson at site $j$, with the hardcore constraints $\hat b^\dagger_j\hat b^\dagger_j=\hat b_j\hat b_j=0$. $x_j=(j-L/2)a$ is the position of site $j$, with $a$ being the lattice spacing. In the limit $n_j=\langle\hat b^\dagger_j \hat b_j\rangle\to 0$ for all sites, the lattice Hamiltonian in Eq.~\eqref{eq:HCB} is equivalent to the continuum one in Eq.~\eqref{eq:H_LL2}. The parameters for the two Hamiltonians are related via $J=\hbar^2/(2ma^2)$~\cite{Rigol2005gpo,Wilson2020ood}.

Since we model the experimental system at finite temperature, the initial density matrix of each 1D tube $l$ is written as
\begin{equation}
    \rho_l(t=0)=\frac{1}{Z_l} \exp\left(-\frac{\hat H^i_l-\mu_l\hat N }{k_B T_l}\right),
\end{equation}
where
\begin{equation}
    Z_l=\text{Tr}\left[\exp\left(-\frac{\hat H^i_l-\mu_l \hat N}{k_B T_l}\right)\right]
\end{equation}
is the partition function, $k_B$ is the Boltzmann constant, $T_l$ is the initial temperature, $\hat{N}$ is the total particle number operator, and $\hat H^i_l$ is the initial Hamiltonian. After the quench, the density matrix evolves in time and is given by
\begin{equation}
    \hat \rho_l(t)=\exp  \left( \! -\frac{i}{\hbar}\hat H^f_l t\right) \rho_l(t=0) \exp \left(\frac{i}{\hbar}\hat H^f_l t\right).
\end{equation}
The one-body density matrix of tube $l$, from which the density and momentum distributions of the hardcore bosons are obtained, is defined as
\begin{equation}
\rho^l_{ij}(t)=\text{Tr}\left[\hat b_i^\dag \hat b^{}_j \hat \rho_l(t) \right] .
\label{eq:ETOBDM_FT}
\end{equation}
We compute $\rho^l_{ij}(t)$ exactly by mapping the hardcore boson model onto spinless fermions and using properties of Slater determinants, following the approach developed in Refs.~\cite{Rigol2005fpo,Xu2017eoo}. The density is given by the diagonal of $\rho^l_{ij}(t)$, while the momentum distribution $f^l(k,t)$ is given by the Fourier transform of $\rho^l_{ij}(t)$. Adding the results for all tubes, we obtain the full distributions compared to the experimental results. For our calculations, we choose $a=3.2\times10^{-8}$~m as the lattice spacing and consider systems up to $L=1500$. 

\begin{figure*}[!t]
\includegraphics[width=0.9\textwidth]{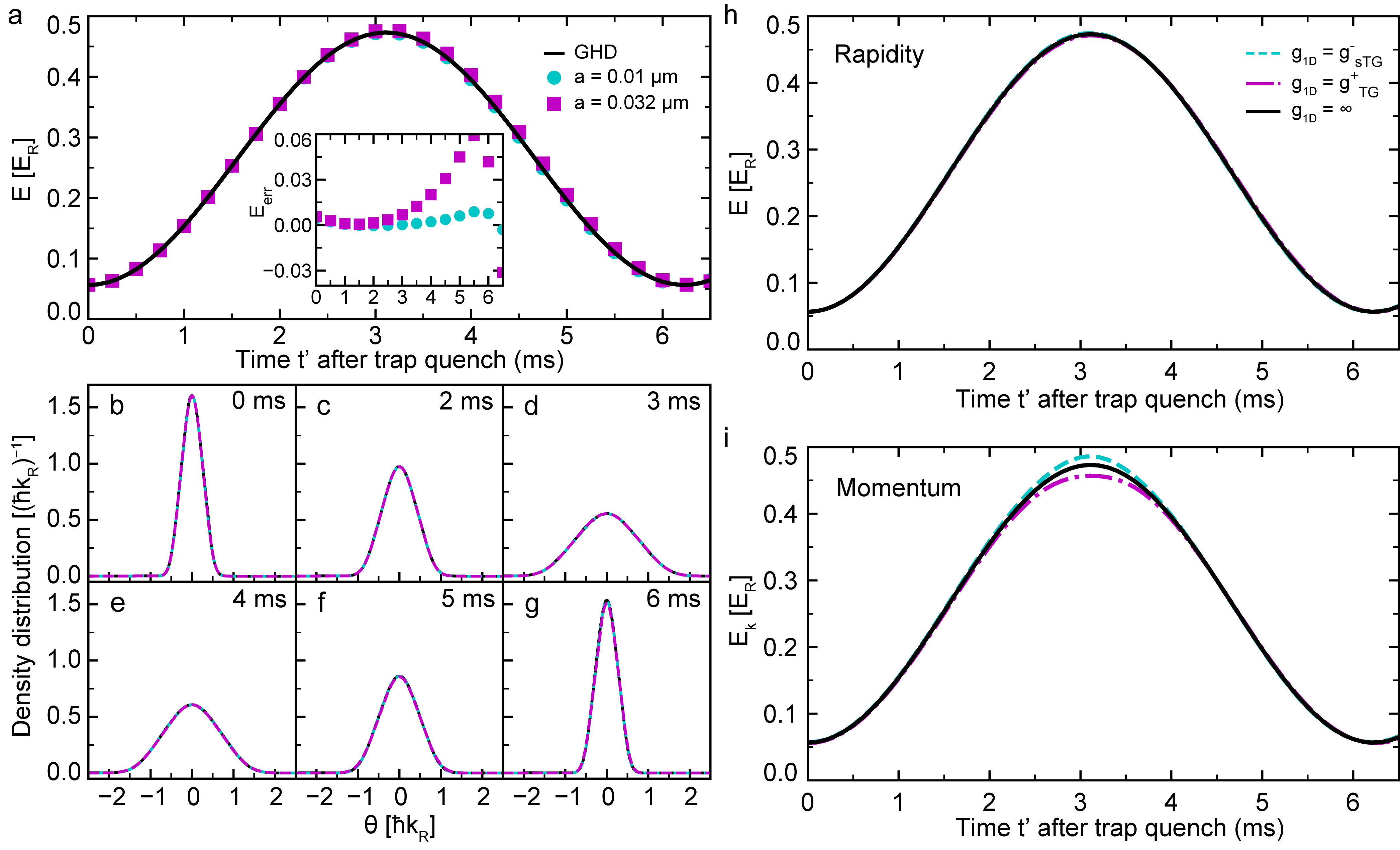}
\caption{Discussions of TG theory. (Left panels) Effect of lattice discretization in the TG calculations. (a) Rapidity energies computed using the lattice calculations with $a=0.01$~$\mu$m (circles) and $a=0.032$~$\mu$m (squares), and using GHD (solid line). Inset: Relative difference $E_\text{err}=|E_\text{latt}-E_\text{GHD}|/|E_\text{GHD}|$ between the energies computed using the lattice discretizations ($E_\text{latt}$) and the GHD results ($E_\text{GHD}$). (b--d) Rapidity distributions from the lattice calculations with $a=0.01$~$\mu$m (dashed cyan lines), $a=0.032$~$\mu$m (dashed-dotted magenta lines), and from GHD (solid lines) at $t'=0$, 2, 3, 4, 5, and 6~ms, resp. (Right panels) Energies from (h) rapidity, and (i) momentum (i) for quenches with large $\gamma$. The results from our GHD calculations are plotted for a repulsive gas with  $g_{\rm 1D}=g^+_\text{TG}=263 \hbar^2/m$~$\mu$m$^{-1}$, an attractive gas with $g_{\rm 1D}=g^-_\text{sTG}=-338 \hbar^2/m$~$\mu$m$^{-1}$, and the exact TG case $g_{\rm 1D}=\infty$ versus time.}
\label{fig:TG_theory}
\end{figure*}

To explore the effect that the finite TOF duration has on the experimental measurements of the momentum distribution functions, we also compute the density distribution $n^{\rm TOF}_j(t;t_{\rm 3D})$ after the free 3D TOF expansion for a time $t_{\rm 3D}$ following the 1D evolution for a time $t$:
\begin{equation}\label{eq:expansion}
n^{\rm TOF}_j(t;t_{\rm 3D})=\sum_{m,n} G^*_{j,m}(t_{\rm 3D})G^{}_{j,n}(t_{\rm 3D})\rho^{}_{m,n}(t)\,,
\end{equation}
where 
\begin{equation}
G_{m,n}(t_{\rm 3D})=\sum_{k}\exp\left(-\frac{it_{\rm 3D}}{\hbar} \left[\epsilon_k-\frac{\hbar k(x_m-x_n)}{t_{\rm 3D}}\right]\right)
\end{equation}
is the free one-particle propagator, with $\epsilon_k=\hbar^2 k^2/2m$ \cite{Rigol2005fia}. $n^{\rm TOF}_j$ is used to determine the TOF-corrected FWHM of the momentum distribution in the context of the 10$\times$ sTG quench experiment reported in Fig.~2(b) in the main text. 

We also find that the rapidity distributions are modified by the finite TOF, albeit generally not as much as the momentum distributions. To compute the rapidity distributions measured after a finite TOF, we assume that the momentum distribution of the bosons has fully fermionized after the 1D expansion. Consequently, the ensuing 3D TOF expansion of the bosonic density is identical to that of noninteracting fermions, which can be efficiently computed numerically using Eq.~\eqref{eq:expansion} with the fermionic (as opposed to the computationally more costly bosonic) one-body density matrix after the 1D expansion. This is how we compute the TOF-corrected FWHM of the rapidity distribution in the context of the 10$\times$ sTG quench experiment reported in Fig.~2(b) in the main text.

In Figs.~\ref{fig:TG_TOF}(a,b), we directly compare our TOF-corrected numerical results (dashed) with experiment data (solid) for (a), momentum and (b), rapidity distributions at different times after the quench. The distributions generally agree, except for the momentum distribution at 3.3~ms, where the experimental distribution is more bosonic at high density. This is an expected consequence of the decrease of the $|\gamma|$ at maximal compression, which is not accounted for in the theory in the TG limit as we discuss below.

Figures~\ref{fig:TG_TOF}(c,d) plot the FWHM of the (c), momentum and (d), rapidity distributions with and without the TOF correction. The finite TOF produces both a time delay and an asymmetry about the maximal compression point, as also noted in Ref.~\cite{Wilson2020ood}. In Figs.~\ref{fig:TG_TOF}(e,f), we show the kinetic and total energies computed from the corresponding momentum and rapidity distributions, resp.  We find only a time shift and no other distortions between the TOF-corrected curves their uncorrected counterparts. This is apparent in Figs.~\ref{fig:TG_TOF}(g,h), where we shift the energies to align with them. 

To align these curve as well as those in Fig.~3, we first find a continuous fitting function by applying linear interpolation to the simulation curve. The only free parameter in the fitting function is the time offset. We then fit the TOF-corrected energies to obtain the optimal time shift for each configuration. This procedure is used in Fig.~3 of the main text to align the GHD results for the kinetic and total energies (which we cannot correct for finite TOF effects) with the experimentally measured ones. Time shifts for each configuration in Fig.~3 are tabulated in Fig.~\ref{fig:Parameters}(d). For momentum data in Figs.~3(d,e), we exclude data points between $t' = 2.3$--4.3~ms from fitting to avoid bias caused by the energy dip. 

The fact that our calculations in the TG limit are carried out at finite temperature strongly limits the lattice sizes that we can solve exactly using the approach in Refs.~\cite{Rigol2005fpo,Xu2017eoo} when compared to zero temperature calculations, such as the ones carried out in, e.g., Refs.~\cite{Rigol2005gpo,Wilson2020ood}. As a result, our finite-temperature calculations suffer from stronger lattice discretization effects. In Fig.~\ref{fig:TG_theory}(a--g), we plot (a), the rapidity energy and (b--g), the rapidity distributions obtained using two different lattice discretizations and GHD. Unlike the momentum distributions, the rapidity distributions and their associated rapidity energies can be exactly computed using much larger lattices (they are the same as for free fermions) and are also accessible with our GHD calculations. For the rapidity energy in Fig.~\ref{fig:TG_theory}(a), one can see that the discretization errors (whose values are reported in the inset) for the lattice discretization used in our calculations of the momentum distributions are smaller than 2\% for $t<4$~ms and less than 1\% at all times for a lattice spacing that is 3$\times$ smaller. The errors, which are relative errors, are largest close to the end of the first oscillation period where the energy attains its minimum value. The results in Figs.~\ref{fig:TG_theory}(b--g) show that the lattice discretization has no visible effect in the calculated rapidity distributions. Hence, we expect that all of our experiment-theory comparisons based on the TG calculations are not affected by the lattice discretization effects.

Another possible source of difference between our theoretical results in the TG regime and the experimental ones is the fact that in the latter $|g|$ is large but finite [see Fig.~\ref{fig:TG_TOF}(a) at 3.3 ms]. In Figs.~\ref{fig:TG_theory}(h,i), we show results for the following cases:  repulsive state with $g_{\rm 1D}=g^+_\text{TG}=263 \hbar^2/m$~$\mu$m$^{-1}$, $g_{\rm 1D}=\infty$, and attractive state with $g_{\rm 1D}=g^-_\text{sTG}=-338 \hbar^2/m$~$\mu$m$^{-1}$. They are simulated with the same distribution of atoms among the tubes. The rapidity energies in Fig.~\ref{fig:TG_theory}(h) are indistinguishable from each other. The kinetic energies in Fig.~\ref{fig:TG_theory}(i) exhibit very small differences close to the maximum compression point, as expected from the generation of interaction energy as the density increases.  This results in a decrease (increase) of the kinetic energy at $g^+_\text{TG}$ ($g^-_\text{sTG}$). 

\end{document}